\documentclass[sigconf]{acmart}

\AtBeginDocument{%
  }

\usepackage{url}
\usepackage{graphicx}
\usepackage{subcaption}
\usepackage{array}
\usepackage{booktabs}
\usepackage{multirow}
\usepackage{caption}
\usepackage{makecell}
\usepackage{geometry}
\usepackage{enumitem}
\usepackage{algorithm}
\usepackage{stfloats}
\usepackage[noend]{algpseudocode}
\usepackage[htt]{hyphenat}

\setcopyright{none} 
\newcommand{\Ryzen}{AMD Ryzen\textsuperscript{\texttrademark} AI}

\setcopyright{none}
\renewcommand\footnotetextcopyrightpermission[1]{}

\begin{document}

\newcommand{\eddie}[1]{\textcolor{orange}{[EDDIE: #1]}}


\title{HeteroMosaic: Exposing and Exploiting Heterogeneous Execution Opportunities for Energy-Efficient Edge LLM Inference}
\titlenote{To appear in the Proceedings of MICRO 2026.}



\settopmatter{authorsperrow=4}

\author{Gregory Hyegang Jun}
\affiliation{%
  \institution{UIUC}
  \country{USA}
}
\email{hgjun2@illinois.edu}

\author{Wesley Pang}
\affiliation{%
  \institution{UIUC}
  \country{USA}
}
\email{qpang2@illinois.edu}

\author{Eddie Richter}
\affiliation{%
    \institution{\mbox{\fontsize{9.5}{10}\selectfont Advanced Micro Devices, Inc.}}
    \country{USA}
}
\email{eddie.richter@amd.com}

\author{Mehdi Saeedi}
\affiliation{%
    \institution{\mbox{\fontsize{9.5}{10}\selectfont Advanced Micro Devices, Inc.}}
    \country{Canada}
}
\email{mehdi.saeedi@amd.com}

\author{Aporva Amarnath}
\affiliation{%
    \institution{\mbox{\fontsize{9.5}{10}\selectfont Advanced Micro Devices, Inc.}}
    \country{USA}
}
\email{aporva.amarnath@amd.com}

\author{Pallavi Ferrao}
\affiliation{%
    \institution{\mbox{\fontsize{9.5}{10}\selectfont Advanced Micro Devices, Inc.}}
    \country{Canada}
}
\email{pallavi.ferrao@amd.com}

\author{Deming Chen}
\affiliation{%
  \institution{UIUC}
  \country{USA}
}
\email{dchen@illinois.edu}

\renewcommand{\shortauthors}{Jun et al.}



\begin{abstract}

Modern edge system-on-chips (SoCs) increasingly combine CPUs, integrated graphics processing units (iGPUs), and neural processing units (NPUs) to meet the growing compute demands of edge AI. However, existing LLM runtimes do not fully optimize this heterogeneity. Production frameworks typically make only coarse device-level decisions, while prior research often focuses on optimizing operations locally rather than holistically accelerating the LLM task graph. As a result, much of the available heterogeneous opportunity remains unrealized, especially on unified-memory edge platforms where performance depends not only on where an operation runs, but also on how execution is coordinated across the task graph.

We argue that edge LLM inference on such systems should be treated as a heterogeneity-first scheduling problem rather than a coarse-grained accelerator-mapping problem, and present HeteroMosaic to realize this view. We first develop a heterogeneous roofline model to characterize the potential benefit of heterogeneity across systems with different relative iGPU and NPU capabilities. Guided by this analysis, HeteroMosaic decomposes LLM execution into dependency-preserving micro-batches that expose additional opportunities to overlap work across accelerators. It then uses trace-guided optimizations to jointly tune this new schedule with device allocation so that critical stages complete earlier despite real system effects such as device variation, unified-memory contention, DVFS, and NPU runtime behavior.

We implement HeteroMosaic in PyTorch C++ and evaluate it on three \Ryzen~platforms spanning NPU-heavy, balanced, and iGPU-heavy designs. On a balanced system, across select off-the-shelf models, HeteroMosaic achieves up to 1.73$\times$ speedup over an iGPU baseline, 1.78$\times$ over an NPU baseline, and up to 2.05$\times$ over strong existing frameworks such as \texttt{llama.cpp}, while using up to 45.3\% less energy. Compared with prior heterogeneous edge AI solutions, HeteroMosaic improves performance by up to 2.35$\times$.

\end{abstract}




\maketitle
\fancyhead{}
\fancyhead[LO]{\footnotesize\shorttitle}
\fancyhead[RO]{\footnotesize\shortauthors}
\fancyhead[LE]{\footnotesize\shortauthors}
\fancyhead[RE]{\footnotesize\shorttitle}
\makeatother

\section{Introduction}

\label{sec:heteromosaic-software-stack}
\begin{figure*}[t]
  \centering
  \includegraphics[width=0.91\linewidth]{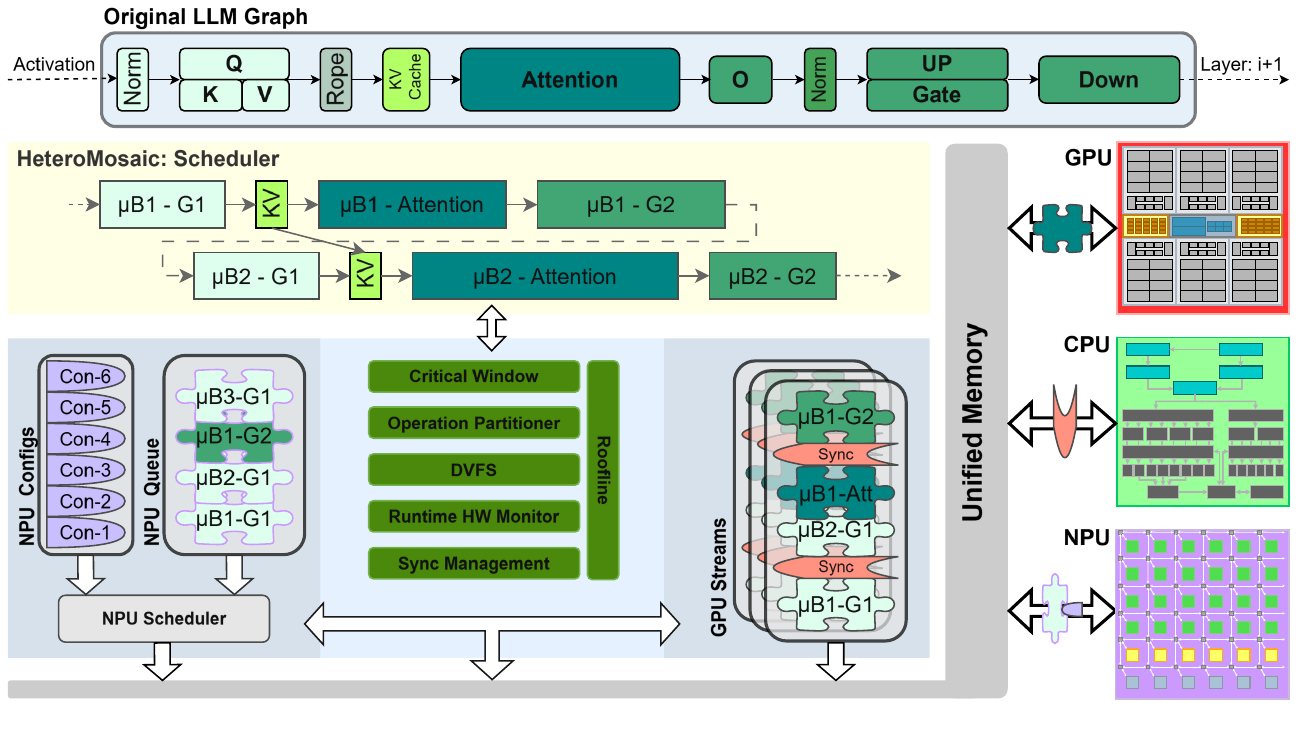}
  \vspace{-20pt}
    \caption{Overview of HeteroMosaic. Using off-the-shelf LLMs, HeteroMosaic restructures execution into dependency-preserving micro-batches and jointly optimizes cross-accelerator overlap and per-stage assignment using a heterogeneous roofline model and trace-guided critical-interval co-optimization. The resulting schedule executes across the CPU, iGPU, and NPU with unified memory, shared weights, low-overhead synchronization, and NPU-aware queue management.}
  \label{fig:hetero-flow}
\end{figure*}

Large language models (LLMs) are transforming computing systems from passive tools into active agents that interpret, generate, and reason over complex data~\cite{openai_chatgpt_2023,openai_gpt4_2023,anthropic_claude_code_2024,openai_khan_academy_2023}. As advances in quantization, distillation, and efficient model design push increasingly capable models from datacenters to user-facing devices, edge inference is becoming attractive not only for privacy and cost, but also for latency-sensitive applications such as personal assistants, AR/VR, robotics, local search, and industrial automation~\cite{siriwardhana2021survey,tahir2025edge,openvla_2024,black2024pi0,hariharan2023realtime,wang2024mememo,wang2025leann,wang2025empowering}. These deployments are especially sensitive to time-to-first-token and sustained throughput because the model often sits inside an interactive control loop rather than serving one-shot requests. As a result, practical edge inference is constrained not only by model size, but also by the growing cost of serving long-context workloads under tight latency and power limits.

Modern edge platforms attempt to meet this demand through heterogeneous system-on-chips (SoCs) that combine CPUs, integrated graphics processing units (iGPUs), and neural processing units (NPUs)~\cite{amdnpu,intelnpu,qualcommnpu,applenpu}. Yet current software stacks still expose this heterogeneity mostly through coarse backend choices rather than fine-grained cross-accelerator workload scheduling. AMD exposes GPU and NPU execution through separate software stacks, ROCm and ONNX Runtime~\cite{ROCm_HIP,amd_oga_2026}, respectively; Apple Core ML~\cite{apple_coreml_2026} allows developers to choose compute units while leaving dispatch to the runtime; and Qualcomm QNN / AI Engine Direct~\cite{qualcomm_qnn_2026} likewise relies on explicit backend targets. On unified-memory edge SoCs such as those mentioned above, this abstraction still leaves performance on the table. Unlike datacenter settings with clear host-device memory separation~\cite{FlexGen, lia}, unified memory eliminates host-device copy overheads and makes fine-grained cross-accelerator mapping practical. But this opportunity is rarely explicit in the original LLM graph: it must first be exposed through careful scheduling, and then mapped onto the right accelerator under tight thermal and power constraints.

The key open question is whether heterogeneity itself improves end-to-end LLM inference once confounding factors are removed. Prior edge LLM systems show that heterogeneous SoC execution can be useful, but their gains are often coupled with model restructuring, sparsity, outlier routing, or quantization changes~\cite{xu2025llmnpu,xue2024powerinfer2}. GPU--NPU systems such as HeteroInfer~\cite{chen2025heteroinfer} study heterogeneous mapping more directly, but still focus primarily on greedy tensor-level placement on highly imbalanced mobile SoCs. More broadly, prior work does not jointly study two complementary sources of heterogeneous opportunity, overlapping independent work across the graph and splitting individual operations across accelerators. This gap matters most on unbalanced SoCs, where a policy that only optimizes local operation placement can either mostly reproduce the stronger accelerator's behavior or offload too much work to the weaker one, obscuring when heterogeneity is genuinely beneficial.

HeteroMosaic addresses this gap by treating edge LLM inference on heterogeneous SoCs as a scheduling problem, not just as an accelerator placement problem. We first develop a heterogeneous roofline model to characterize when heterogeneous execution should help across systems with different relative iGPU and NPU capabilities. Guided by this analysis, HeteroMosaic decomposes execution into dependency-preserving micro-batches that expose additional opportunities to overlap work across accelerators. Exposing this overlap is only the first step: once these new scheduling choices exist, their benefits become interdependent on real systems. On real systems, the best greedy accelerator assignment is not always the best global choice. A mapping that is faster in isolation can still degrade end-to-end performance if, because of accelerator asymmetry, unified-memory contention, DVFS, or NPU runtime behavior, it delays stages that are more critical to the overall execution graph. HeteroMosaic therefore uses trace-guided critical-interval co-optimization to jointly tune the schedule, decide which accelerator executes each stage, and adjust stage completion times around the measured critical path.

We realize HeteroMosaic on three \Ryzen~ SoCs spanning NPU-heavy, balanced, and iGPU-heavy designs. We focus on \Ryzen~ for two reasons. First, \Ryzen~ exposes low-level software stacks for both the iGPU and NPU, enabling optimized custom kernels on each accelerator together with low-overhead interoperability and control across them. Second, \Ryzen~ offers a broad range of CPU, iGPU, and memory configurations, allowing us to study heterogeneous execution across diverse balance points while keeping the LLM model and framework fixed. Across these systems, we show that heterogeneity can be the winning solution when it is jointly scheduled at the graph and operation levels. Our key contributions are as follows.

\begin{enumerate}[leftmargin=1.5em]
    \item[$\bullet$] \textbf{Problem reformulation}. We show that heterogeneous edge LLM inference on unified-memory SoCs is fundamentally a \emph{scheduling} problem, not just an operation-placement problem. Much of the heterogeneous opportunity is not explicit in the original execution graph and must first be exposed before it can be exploited.

    \item[$\bullet$] \textbf{Analytical model}. We develop a heterogeneous roofline model that characterizes when heterogeneous execution should help across systems with different relative iGPU and NPU capabilities, and use it as a principled target for the runtime.

    \item[$\bullet$] \textbf{Algorithmic}. We show that decomposing LLM execution into dependency-preserving micro-batches exposes additional opportunities to overlap work across accelerators that are unavailable in the monolithic schedule.

    \item[$\bullet$] \textbf{Scheduling}. We show that exploiting this exposed opportunity is inherently non-greedy on real systems, and introduce trace-guided critical-interval co-optimization to jointly tune the schedule, accelerator assignment, and stage timing under practical runtime effects such as DVFS, device asymmetry, and NPU runtime behavior.

    \item[$\bullet$] \textbf{Systems and empirical}. We realize this design on three \Ryzen~ SoCs spanning NPU-heavy, balanced, and iGPU-heavy designs, and show that heterogeneity-first scheduling outperforms strong single-accelerator and prior heterogeneous baselines while reducing energy and improving tokens per watt by up to 45.3\%. Compared with prior SOTA, we achieve an average 1.25× speedup across platforms and models, with a peak of 2.35× and a minimum of 1.07×.
\end{enumerate}

\section{Background}

\begin{table}[t]
\centering
\caption{\Ryzen~ SoC specifications. All CPUs use Zen~5 cores. The table reports CPU core/thread count, iGPU RDNA generation and compute units (CUs), NPU generation and AIE core count, and peak memory bandwidth under the maximum LPDDR5X configuration.}
\label{tab:ryzen-socs}

\setlength\tabcolsep{4pt}
\scalebox{0.78}{
\begin{tabular}{llllll}
\toprule
\textbf{SoC} &
\makecell[c]{\textbf{CPU}\\\textbf{Zen~5}} &
\makecell[c]{\textbf{iGPU}\\\textbf{RDNA / CUs}} &
\makecell[c]{\textbf{NPU}\\\textbf{Gen.}} &
\makecell[c]{\textbf{NPU}\\\textbf{AIEs}} &
\makecell[c]{\textbf{Mem.}\\\textbf{BW}} \\
\midrule
\makecell[c]{AMD Ryzen\textsuperscript{\texttrademark}\\ AI 7 350}     & 8c/16t   & RDNA 3.5, 8 CUs  & XDNA 2 & 32 & 128 GB/s \\
\midrule
\makecell[c]{AMD Ryzen\textsuperscript{\texttrademark}\\ AI 9 HX 370}  & 12c/24t  & RDNA 3.5, 16 CUs & XDNA 2 & 32 & 128 GB/s \\
\midrule
\makecell[c]{AMD Ryzen\textsuperscript{\texttrademark}\\ AI Max+ 395}  & 16c/32t  & RDNA 3.5, 40 CUs & XDNA 2 & 32 & 256 GB/s \\
\bottomrule
\end{tabular}
}
\end{table}

\begin{figure*}[!t]
  \centering

  \begin{subfigure}[b]{0.31\linewidth}
    \centering
    \includegraphics[width=0.90\linewidth ]{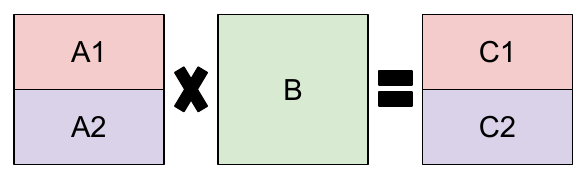}
    \caption{M Split}
    \label{fig:partitionM}
  \end{subfigure}
  \hspace{0.01\linewidth}
  \vrule width 0.5pt
  \hspace{0.01\linewidth}
  \begin{subfigure}[b]{0.31\linewidth}
    \centering
    \includegraphics[width=0.90\linewidth]{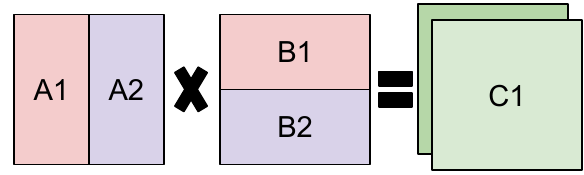}
    \caption{K Split}
    \label{fig:partitionK}
  \end{subfigure}
  \hspace{0.01\linewidth}
  \vrule width 0.5pt
  \hspace{0.01\linewidth}
  \begin{subfigure}[b]{0.31\linewidth}
    \centering
    \includegraphics[width=0.90\linewidth]{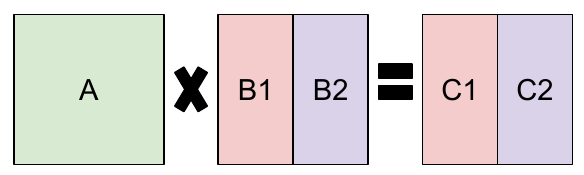}
    \caption{N Split}
    \label{fig:partitionN}
  \end{subfigure}

  \caption{Different tensor partitioning schemes in GEMM.}
  \label{fig:gemmpartition}
\end{figure*}

\subsection{\Ryzen}

\Ryzen~ SoCs~\cite{amd_ryzen_ai_7_350_2026, amd-2025-ryzen-ai-9-hx-370, amd_ryzen_ai_max_plus_395_2026} contain an x86 CPU, an RDNA iGPU, and an XDNA NPU. These components are connected through AMD's Infinity Fabric, which integrates the compute engines and connects them to a unified memory architecture. The \Ryzen~ product family spans multiple use cases and system requirements. Table~\ref{tab:ryzen-socs} summarizes three current generation \Ryzen~ configurations, showing differences in CPU resources, iGPU size, NPU generation, and memory bandwidth.

\subsection{LLM Inference}
LLM inference consists of two distinct phases. The first phase, \textit{prompt encoding} or \textit{prefill}, processes the entire prompt and produces the first output token. Prefill is dominated by large matrix multiplications and is therefore primarily constrained by dense general matrix multiply (GEMM) throughput. The commonly used metric \textit{time to first token} (TTFT) captures the latency of processing a prompt of $N$ tokens and producing the first generated token.

The second phase, \textit{token generation} or \textit{decode}, produces one token at a time by attending to both the encoded prompt and previously generated tokens. Decode relies more heavily on general matrix-vector multiplication (GEMV) operations and KV-cache traffic, and is therefore often constrained by memory bandwidth rather than raw arithmetic throughput. The standard performance metric for this phase is \textit{tokens per second} (TPS), which measures the sustained generation throughput after prefill.

\subsection{LLM Micro-Batching}
Micro-batching~\cite{agrawal2024sarathi,llamacpp_common_2026,llamacpp_discussion_6328} reduces peak prefill memory by splitting a single long prompt into smaller contiguous prompt segments. Here, a micro-batch does not refer to a batch of independent requests, as in conventional training or batched inference. Instead, it refers to a sub-sequence of tokens from the same prompt. Rather than executing one monolithic prefill pass over the full prompt, the runtime processes these prompt segments sequentially. As each micro-batch is processed, the KV cache is incrementally extended, while temporary activations are materialized only for the current segment.

Importantly, attention can still attend to preceding tokens through the KV cache, meaning the KV cache is used not only during decode but also during prefill. Micro-batch size determines the size of $Q$, while $K$ and $V$ continue to grow through the KV cache. Because LLM models employ causal attention, later micro-batches see a deeper KV history and therefore incur higher attention cost than earlier micro-batches. Causality is what makes this decomposition exact. A later micro-batch may attend to tokens from earlier micro-batches through the KV cache, but it does not depend on tokens that have not yet been processed.

From a computational perspective, micro-batching does not change the final mathematical result, but it does change the execution schedule. It increases dispatch frequency, reuses weights across multiple smaller launches, and changes how attention cost evolves across micro-batches. Those properties make it useful not only as a memory optimization, but also as a scheduling primitive for heterogeneous execution.

\subsection{Tensor Parallelism}
\label{sec:tensorparallel}
Tensor parallelism distributes a single \emph{operation} across multiple compute units~\cite{shoeybi2019megatronlm}. In LLM inference, the most common target \emph{operations} are GEMM operations because they dominate the arithmetic cost of prefill. These dense linear operations are natural split targets because partitioning preserves regular independent subproblems with predictable communication. Figure~\ref{fig:gemmpartition} shows three natural split dimensions. $M$- and $N$-splits partition the input / output space and require gathering of results, while a $K$-split produces partial sums that must be reduced. In unified-memory SoCs, these communication patterns still matter even though the gather or reduction ultimately resolves through unified memory rather than explicit multi-accelerator collectives.

\subsection{Community-Driven NPU Programming}
\label{sec:fusedNPUkernel}
IRON~\cite{hunhoff2025efficiencyexpressivityextensibilityclosetometal} is an open-source framework for programming AMD's XDNA NPUs, combining fine-grained control over compute kernels with higher-level abstractions for data movement and synchronization. Its open-source nature has enabled the broader community to build optimized custom kernels on top of this stack. In particular, we leverage community-developed AWQ~\cite{lin2023awq} W4A16 kernels~\cite{glassescrab_mlir_aie_2026} as a key component in enabling heterogeneous execution across the NPU and iGPU.

\section{Related Work}
\label{sec:related}

Prior work on edge LLM inference on unified-memory systems has shown evidence of the potential benefits of heterogeneous SoC execution, but often conflates heterogeneity with other sources of improvement. Existing systems combine heterogeneous execution with model restructuring, quantization changes, activation sparsity, outlier routing, or accelerator capability imbalance, where practical compute throughput is heavily skewed toward one accelerator. Moreover, the closest GPU--NPU edge LLM studies are largely evaluated on Qualcomm Gen3-class mobile SoCs~\cite{qualcomm_snapdragon_8_gen3_2023}, where HeteroInfer reports a roughly $1{:}10$ practical GPU-to-NPU throughput ratio~\cite{chen2025heteroinfer}. As a result, it is difficult to determine whether the reported gains come from heterogeneity itself or from coupled factors such as reduced work, changed numerical formats, platform-specific quantization paths, or accelerator imbalance. Thus, while prior work establishes that SoC heterogeneity can be useful, it leaves open the key question: can heterogeneity itself, when optimized from first principles, become the winning strategy for off-the-shelf LLM architectures and standard quantization formats?

\subsection{Sparse and Model-Restructured Inference.}

\subsubsection{Activation-sparse and storage-aware inference.} PowerInfer-2~\cite{xue2024powerinfer2} improves edge LLM inference by combining heterogeneous execution with activation sparsity and storage-aware execution. Rather than executing each linear layer as a dense operation, it decomposes computation into fine-grained neuron clusters, maps dense activation clusters to the NPU, and handles sparse clusters on the CPU. It further coordinates compute with flash I/O through segmented neuron caching and pipelining, allowing models that exceed DRAM capacity to execute more efficiently.

The key distinction is that PowerInfer-2 changes the amount and structure of computation exposed to the hardware. Activation sparsity allows inactive or low-importance neuron clusters to be skipped or delayed, so part of the speedup may come from reducing the effective number of dense operations rather than from better scheduling the same dense workload across accelerators. This does not diminish the value of PowerInfer-2, but it makes the source of improvement different from HeteroMosaic. HeteroMosaic keeps the off-the-shelf LLM graph and standard weight-only quantization path intact, then improves performance by changing when and where existing work executes.

In addition, sparsity policies are only as reliable as the calibration or profiling inputs used to construct them. If those inputs do not match deployment prompts, the active-neuron distribution can shift, causing the sparse execution plan to lose either performance predictability or accuracy.

\subsubsection{Outlier-routed NPU inference.}
Fast On-device LLM Inference with NPUs~\cite{xu2025llmnpu} leverages heterogeneity by adapting the model execution path to better match the target hardware. During quantization, the original dense matrix computation is split into two paths: a low-precision dense main path and a high-precision shadow-outlier path for activation--weight interactions that cannot be safely represented within the low-precision format. At runtime, the low-precision main path is mapped to the NPU, while the high-precision outlier path is executed on the CPU. In this sense, the system exploits heterogeneity by reshaping the model so that each backend handles the precision and compute intensity it supports best. This work is among the first to study outlier execution from a hardware-systems perspective, while later work such as SVDQuant~\cite{li2024svdquant} studies the quantization side of outlier handling more directly.

Although this work makes NPU execution practical under mobile-SoC constraints, it also couples heterogeneity with a changed execution structure. The model is explicitly restructured so that the main low-precision path fits the NPU, while the numerically sensitive outlier path is separated onto the CPU. As a result, the reported benefit reflects both hardware-aware outlier decomposition and heterogeneous CPU--NPU mapping. This differs from HeteroMosaic, which does not introduce a separate numerical path or depend on outlier routing; instead, it keeps the off-the-shelf LLM execution structure intact and improves performance by exposing graph-level overlap, shaping stage latency, and coordinating existing work across accelerators.

Consequently, this model-restructuring requirement~\cite{xu2025llmnpu} narrows the generality of the reported heterogeneous-execution result: because the method relies on separating the computation into an NPU-friendly low-precision main path and a CPU-routed shadow-outlier path, its benefit is tied to how cleanly that decomposition matches the target model and backend capabilities. If the computation does not separate in this way, or if the outlier path becomes too expensive, it is unclear whether the same heterogeneous strategy would remain effective. Moreover, as with activation sparsity, shadow-outlier selection can depend strongly on calibration representativeness: if the calibration data does not capture the activation outliers that appear during deployment, either accuracy can degrade or more work may be routed through the expensive shadow path.

\subsection{NPU Optimizations, Runtime Optimizations, and Tensor Parallelism}

\subsubsection{NPU-oriented optimizations.}
ScalingNPU~\cite{scalingNPU}, although not a heterogeneous inference framework, makes evident the potential of mobile NPUs for efficient LLM execution when the runtime and kernels are co-designed around the hardware. Its key contribution is NPU-centric execution: it uses test-time scaling to expose additional decoding parallelism, then realizes that parallelism through hardware-tailored quantization layouts, fused NPU kernels, and carefully optimized runtime scheduling. In this sense, ScalingNPU shows that NPUs can serve as effective LLM execution engines, but only when the workload structure, data layout, and runtime are designed around the hardware rather than treated as a generic accelerator backend.

ScalingNPU thus demonstrates that practical NPU performance depends heavily on low-level runtime and kernel co-design. Specifically, it shows that Qualcomm Gen3 NPUs cannot efficiently consume the fine-grained weight-only quantization layouts commonly used in LLM deployment, such as AWQ-style W4A16 group quantization. From an accuracy standpoint, ScalingNPU further shows that QNN~\cite{qualcomm_qnn_2026} per-channel W4A16 severely degrades Llama3.2-1B-Instruct accuracy compared with AutoAWQ per-group W4A16, especially on reasoning benchmarks. To address this mismatch, ScalingNPU introduces an NPU-tailored tile quantization layout, fused NPU kernels, and LUT-based implementations that reduce runtime dequantization overhead.

However, optimizing specifically for the NPU can also reduce compatibility with GPU execution paths, since an NPU-tailored weight layout may not be directly usable by GPU kernels. HeteroMosaic targets this broader systems problem by building compatible iGPU and NPU execution kernels, then applying graph-level and operation-level heterogeneity. This is possible because the NPU and GPU execution paths use the same underlying LLM model and quantization scheme, enabling heterogeneous scheduling without requiring separate model representations for each accelerator.

\subsubsection{Heterogeneous Tensor Parallelism.}
HeteroInfer~\cite{chen2025heteroinfer} studies heterogeneity more directly by mapping LLM inference across the GPU and NPU of a mobile SoC. It uses the NPU as the primary compute engine and the GPU as a secondary accelerator, combining layer-level and tensor-level partitioning with profiling-guided scheduling and lightweight synchronization. Its focus is therefore closer to heterogeneous accelerator mapping than sparsity- or outlier-based systems.

At the same time, the evaluation leaves the source of the reported gains difficult to isolate. The evaluated platform is highly NPU-skewed: HeteroInfer reports roughly a $1{:}10$ practical GPU-to-NPU throughput ratio on Snapdragon 8 Gen 3~\cite{qualcomm_snapdragon_8_gen3_2023}. In other words, the NPU is approximately an order of magnitude faster than the GPU for the evaluated kernels. In this regime, assigning more work to the NPU should improve performance even without demonstrating complementary GPU--NPU execution. Thus, without an NPU-only baseline for the same model and software stack, it is difficult to determine whether the reported gains come from heterogeneity itself, from improved tensor placement, or simply from assigning more work to the stronger accelerator. This ambiguity is reinforced by Qualcomm AI Hub results, which report an optimized Llama-v2-7B-Chat deployment for Snapdragon 8 Gen 3 using W4A16 weights with W8A16 in some layers~\cite{qualcomm_llama_v2_7b_chat_2026} that meets or exceeds the heterogeneous performance numbers reported by HeteroInfer.

There are also quantization and backend-equivalence issues that complicate attribution. Related work on Qualcomm Gen3 NPUs, such as ScalingNPU~\cite{scalingNPU}, notes that the available QNN~\cite{qualcomm_qnn_2026} quantization path differs from blockwise LLM quantization schemes and can affect prediction metrics such as perplexity. Since HeteroInfer uses different backend stacks for different devices, QNN for the NPU and OpenCL for the GPU, the evaluation may conflate heterogeneous scheduling with backend-specific quantization and kernel behavior. These factors do not diminish HeteroInfer's contribution as an important GPU--NPU study, but they limit how directly its results establish heterogeneity itself as the source of the gain. HeteroMosaic differs by treating heterogeneity as an end-to-end scheduling problem: it first exposes graph-level overlap through causal micro-batching, then uses heterogeneous partitioning to shape critical and non-critical stage latencies under measured system effects.

\subsubsection{Portable GPU-oriented LLM runtimes.}
\texttt{llama.cpp}~\cite{llama.cpp} is one of the most practical and portable single-accelerator frameworks for edge LLM inference, primarily targeting CPU or GPU execution, with more recent support for select NPU backends. Built around the \texttt{ggml}/GGUF~\cite{ggml_github_2026} execution stack, \texttt{llama.cpp} emphasizes portability across a wide range of backend hardware rather than fine-grained heterogeneous orchestration across accelerators. Its backend ecosystem spans Apple Silicon through Metal~\cite{apple_metal_2026}, NVIDIA GPUs through CUDA~\cite{nvidia_cuda_docs_2026}, AMD GPUs through HIP/ROCm~\cite{rocm}, Intel and NVIDIA GPUs through SYCL~\cite{khronos_sycl_2026,llamacpp_build_2026}, generic GPUs through Vulkan and OpenCL~\cite{khronos_vulkan_2026,khronos_opencl_2026}, CPUs through BLAS/BLIS and ZenDNN~\cite{netlib_blas_2026,blis_github_2026,amd_zendnn_2026}, and emerging accelerator targets such as OpenVINO, CANN, Hexagon, WebGPU, RPC, and VirtGPU~\cite{intel_openvino_2026,huawei_cann_2026,qualcomm_hexagon_sdk_2026,w3c_webgpu_2026,llamacpp_rpc_2026,llamacpp_build_2026}. This broad backend support makes \texttt{llama.cpp} a strong community-driven deployment framework, but its primary abstraction is backend portability rather than heterogeneous graph scheduling.

\texttt{llama.cpp} also provides its own GGUF quantization formats to balance model size, bandwidth, and performance across devices, including 4-bit, 5-bit, 6-bit, 8-bit, and 16-bit weight formats~\cite{llamacpp_quantize_2026}. It also supports micro-batching to reduce peak memory pressure on low-memory systems by splitting long prompt processing into smaller execution units~\cite{llamacpp_common_2026,llamacpp_discussion_6328}. However, in \texttt{llama.cpp}, micro-batching is primarily a memory-management mechanism within a selected backend path, not a mechanism for exposing cross-accelerator overlap. Moreover, because \texttt{llama.cpp} prioritizes portability across diverse backends, its execution path can leave performance on the table on GPUs; for example, backend-agnostic tensor transitions may introduce unnecessary data movement or intermediate casting to preserve a common execution model across hardware targets. HeteroMosaic targets a different point in the design space: it extends micro-batching into a heterogeneous scheduling primitive, builds compatible iGPU and NPU kernels around a shared quantization path, and then combines graph-level overlap with operation-level partitioning to improve end-to-end latency and energy.

\section{Heterogeneous Roofline Model}
\label{sec:roofline}

\begin{figure}[t]
  \centering
    \includegraphics[width=\linewidth]{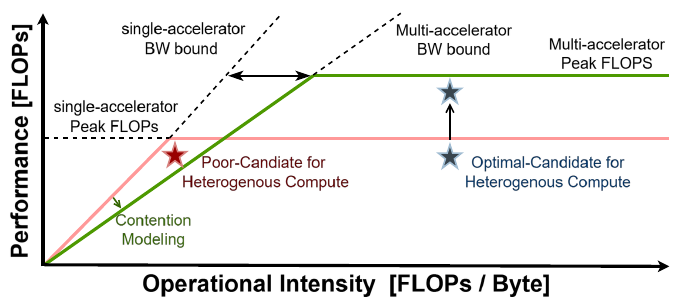}
  \caption{Illustration of the heterogeneous roofline model. Some operations are strong candidates for heterogeneous execution, while others remain bandwidth-bound when viewed in isolation.}
  \label{fig:roofline}
\end{figure}

We develop an operation-level heterogeneous roofline model for our target \Ryzen~ platforms, illustrated in Figure~\ref{fig:roofline}. The figure first contrasts two bounds: the red curve denotes the conventional single-accelerator roofline, while the green curve denotes the best-case operation-level heterogeneous roofline when multiple accelerators are active. The red curve is limited by the compute throughput and memory bandwidth of a single accelerator. The green curve raises the peak operation throughput as heterogeneous execution combines compute resources across accelerators. At the same time, both curves remain constrained by the unified off-chip memory system and by system effects such as DVFS.

Figure~\ref{fig:roofline} should therefore be interpreted as an operation-level view of attainable performance, not as a roofline for the full end-to-end LLM workload. The blue point denotes an operation that is directly suitable for heterogeneous execution, since, relative to the single-accelerator roofline, it remains sufficiently compute-limited that heterogeneous execution can raise its attainable performance toward the heterogeneous bound. The red point denotes an operation that is a poor heterogeneous candidate because of its proximity to the bandwidth-bound regime and therefore cannot benefit much from additional heterogeneous compute capability.

The roofline model therefore serves two purposes. First, it quantifies the best-case upper bound of heterogeneous opportunity across different relative accelerator capabilities within a SoC. Second, it reveals why a placement-only view is insufficient. Some operations are good heterogeneous targets in the original graph, while others appear unsuitable under local operation placement alone. However, these seemingly unsuitable opportunities can become useful when scheduling moves bandwidth-bound work off the critical path or exposes new overlap that was not visible in the original monolithic graph.

This distinction explains why an operation-level roofline is necessary but not sufficient for end-to-end LLM inference. Although it can identify which individual operations are promising heterogeneous candidates, it cannot determine whether accelerating a given operation shortens the global application-level schedule, nor can it expose overlap that is absent from the original execution graph. We therefore treat the roofline as an analytical target rather than an end-to-end predictor. The scheduling framework in the following section restructures execution through causal micro-batching and uses trace-guided critical-interval optimization to determine how much of that target can be realized in practice.

Having established the operation-level intuition behind Figure~\ref{fig:roofline}, we now formalize the heterogeneous roofline. We start from the classical roofline model and extend it to \Ryzen~ platforms that combine CPU, iGPU, and NPU resources under a shared memory system. The classical roofline model~\cite{Williams2009Roofline} defines single-accelerator attainable performance as
\[
P = \min(P_{\text{peak}}, B \cdot I),
\]
where $P_{\text{peak}}$ is peak compute throughput, $B$ is peak off-chip memory bandwidth, and $I$ is operational intensity, defined as arithmetic operations per byte accessed. For floating-point kernels, this corresponds to FLOPs/byte; for quantized kernels, we report ops/byte to match the TOP/s-based throughput model. This model captures the trade-off between compute-bound and memory-bound regimes, which we adapt to reason about heterogeneous execution across multiple accelerators sharing a common memory system.

We adopt Gables~\cite{Gables} to model $N$ compute units operating concurrently on a unified memory system. Each accelerator $i$ has peak compute throughput $P_{\text{peak},i}$, memory bandwidth $B_i$, and executes a fraction $f_i$ of the workload with operational intensity $I_i$, where $\sum_i f_i = 1$. We write the attainable whole-operation performance allowed by accelerator $i$ as
\[
P^{i}_{\text{attainable}} =
\frac{\min(B_i \cdot I_i,\; P_{\text{peak},i})}{f_i}.
\]
The numerator is the local roofline limit for accelerator $i$ on its assigned partition, while the division by $f_i$ converts this local partition throughput into an equivalent whole-operation throughput. The workload fractions $f_i$ correspond to the accelerator-side partitions induced by the heterogeneous tensor split, such as the $M$-, $K$-, and $N$-split GEMM decompositions shown in Figure~\ref{fig:gemmpartition} and described in Section~\ref{sec:tensorparallel}.

For this fixed accelerator-side split, the system-wide average operational intensity, denoted as $I_{\text{avg}} = 1 / \left( \sum_{i=0}^{N-1} \frac{f_i}{I_i} \right)$, is computed using the \textit{weighted harmonic mean} of the individual intensities. For a fixed work split, the attainable heterogeneous performance is limited by the slowest assigned accelerator partition and by the shared-memory bandwidth cap:
\[
P_{\text{attainable}} =
\min\left(
P^{\text{CPU}}_{\text{attainable}},\;
P^{\text{iGPU}}_{\text{attainable}},\;
P^{\text{NPU}}_{\text{attainable}},\;
B_{\text{peak}} \cdot I_{\text{avg}}
\right).
\]
This minimum captures the bottleneck for one heterogeneous split. The full operation cannot complete faster than the slowest participating accelerator or the shared-memory system that feeds them.

The heterogeneous roofline shown as the green curve in Figure~\ref{fig:roofline} is then the best attainable envelope over feasible work splits:
\[
P^{*}_{\text{hetero}} =
\max_{\{f_i\}:\sum_i f_i = 1} P_{\text{attainable}}.
\]
Thus, the inner minimization gives the attainable performance for one split, while the outer maximization selects the split that best balances the CPU, iGPU, and NPU and produces the best-case heterogeneous bound.

To account for practical inefficiencies, we introduce a performance efficiency parameter $\eta_i$ for accelerator $i$, where $0 < \eta_i \leq 1$, and a bandwidth efficiency factor $\alpha_i$. Here, $\eta_i$ captures effective compute efficiency, measured as achieved TOP/s divided by published peak TOP/s, while $\alpha_i$ captures effective memory-bandwidth efficiency after cache behavior, access-pattern effects, unified-memory contention, and runtime overheads. These values are not fixed constants during execution: on real SoCs, they vary with DVFS, thermal state, kernel shape sensitivity, NPU runtime behavior, and memory contention. The effective compute and bandwidth for each accelerator are
\[
P^{\text{eff}}_i = \eta_i \cdot P_{\text{peak},i}, \qquad
B^{\text{eff}}_i = \alpha_i \cdot B_i.
\]
Using these effective terms, the attainable whole-operation performance allowed by accelerator $i$ becomes
\[
P^{i}_{\text{attainable}} =
\frac{\min(B^{\text{eff}}_i \cdot I_i,\; P^{\text{eff}}_i)}{f_i}.
\]
This formulation preserves the roofline structure while allowing the model to account for software overheads, thermal throttling, runtime inefficiencies, and memory contention by scaling effective compute and bandwidth. In particular, when the iGPU and NPU contend for unified memory, effective bandwidth degrades, reducing the attainable performance of both devices and shifting the practical heterogeneous bound downward.

\begin{figure}[t]
  \captionsetup[subfigure]{skip=2pt}
  \centering

  \newcommand{\plotw}{\linewidth}
  \newcommand{\trimplot}[1]{%
    \scalebox{1}[0.80]{%
        \raisebox{-0.5\height}{%
          \includegraphics[width=\plotw,trim=4 8 4 8,clip]{#1}
        }
    }
  }

  \begin{subfigure}[b]{\linewidth}
    \centering
    \trimplot{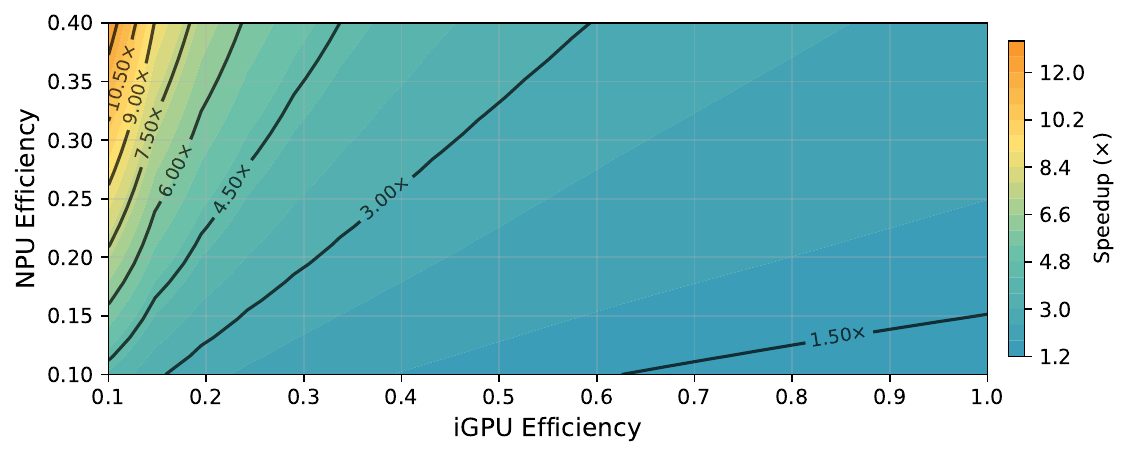}
    \caption{\Ryzen~ 7 350}
    \label{fig:krakn_roofline}
  \end{subfigure}

  \begin{subfigure}[b]{\linewidth}
    \centering
    \trimplot{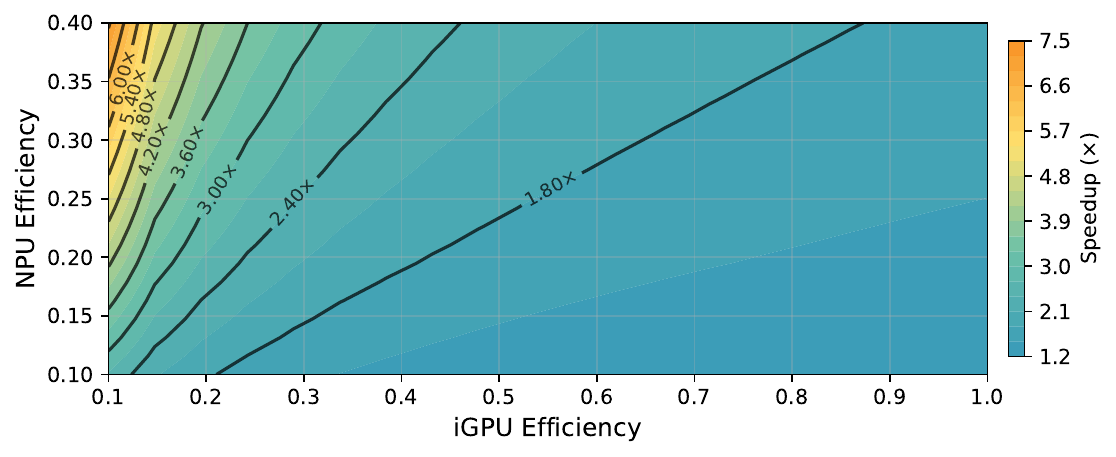}
    \caption{\Ryzen~ 9 HX 370}
    \label{fig:strixP_roofline}
  \end{subfigure}

  \begin{subfigure}[b]{\linewidth}
    \centering
    \trimplot{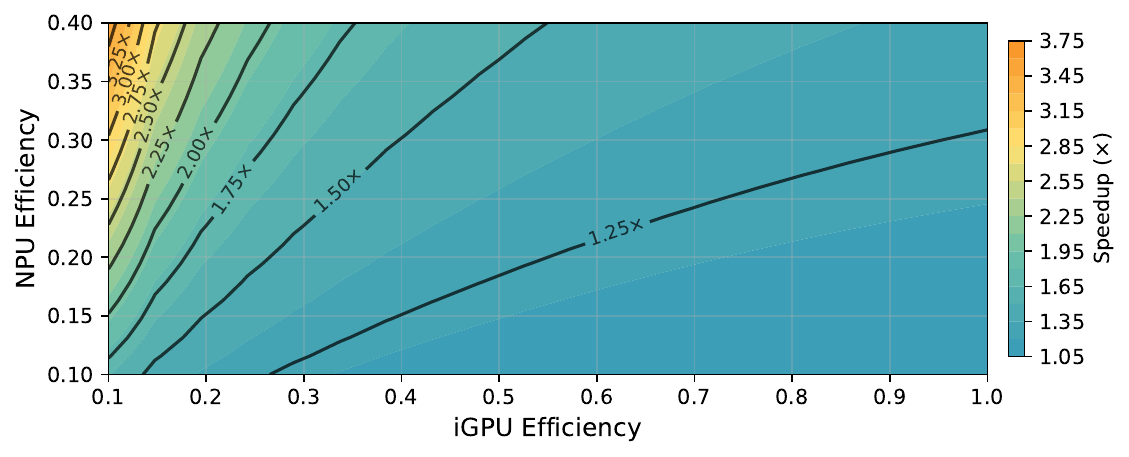}
    \caption{\Ryzen~ Max+ 395}
    \label{fig:strixH_roofline}
  \end{subfigure}
  \caption{Projected heterogeneous speedup over the iGPU baseline from the heterogeneous roofline model as a function of iGPU and NPU efficiency. The contours indicate the projected speedup for different iGPU–NPU efficiencies.}
  \label{fig:roofline_projec}
\end{figure}

We then use this model to project the best-case heterogeneous speedup as a function of achieved iGPU and NPU efficiency. Figure~\ref{fig:roofline_projec} plots this projected speedup surface.\footnote{The CPU is omitted because of its relatively limited dense-compute capabilities.} The figure should not be read as assigning a single fixed $\eta$ or $\alpha$ to each device. Instead, it sweeps achieved iGPU and NPU efficiency values relative to each platform's published peak capability~\cite{amd_ryzen_ai_7_350_2026,amd-2025-ryzen-ai-9-hx-370,amd_ryzen_ai_max_plus_395_2026}. In the idealized projection, runtime overheads and memory-contention losses are excluded; the measured microbenchmarks and traces later calibrate how far real execution falls below this ceiling.

To ground the efficiency ranges used in Figure~\ref{fig:roofline_projec}, we calibrate the NPU and iGPU terms separately rather than assuming that published peak TOPS are fully attainable. For the NPU, we conservatively use 0.4 as a favorable upper efficiency setting, consistent with prior academic studies of AMD NPUs~\cite{wang2025asymmetrictilebufferingbeneficial,FromLoopNeststoSilicon, taka2025strikingbalancegemmperformance}. For the iGPU, prior AMD GPU GEMM benchmarking reports optimized dense-linear-algebra efficiency of roughly 0.7~\cite{gpugemmbenchmarking}. However, we still leave iGPU efficiency as a broader sweep because iGPU performance is more sensitive to operation shape, kernel implementation, and runtime state than isolated GEMM benchmarks. Under the idealized assumption that all compute is heterogeneously mappable, and that software overheads, memory contention, and other runtime effects are absent, the model projects approximate upper bounds of $3\times$, $1.8\times$, and $1.25\times$ over each iGPU baseline for the \Ryzen~ 7 350, \Ryzen~ 9 HX 370, and \Ryzen~ Max+ 395, respectively.

These projected bounds indicate that heterogeneity can be beneficial in principle, but they do not by themselves explain how to realize that opportunity on a real system. In particular, the roofline model does not determine which parts of the graph are immediately profitable to map heterogeneously, which parts become profitable only after schedulable opportunity has been exposed, or how either should be scheduled under practical effects such as unified-memory contention, runtime overheads, and device-specific behavior. The remainder of the paper focuses on exactly this gap: exposing latent heterogeneous opportunity in the execution graph and developing a runtime that can realize as much of the analytical bound as possible in practice.

\section{HeteroMosaic Overview}
Figure~\ref{fig:hetero-flow} illustrates the overall design of HeteroMosaic. HeteroMosaic starts from the fixed structure of off-the-shelf decoder-only LLMs and implements model execution as a compiled C++ function-call path following the transformer architecture. PyTorch C++~\cite{pytorch} is used for tensor and weight management, while the execution graph seen by the scheduler is explicitly exposed through C++ stage boundaries in this custom C++ transformer path. These boundaries form coarse graph nodes, such as $G_1$, Attention, and $G_2$, with edges derived from program order and KV-cache causality. KV-cache updates, RoPE variants, and FlashAttention-style execution are represented as explicit C++/HIP stages, with the FlashAttention-style path implemented as a custom HIP kernel. HeteroMosaic then restructures this explicitly defined graph into causally valid micro-batches and uses the heterogeneous roofline model, microbenchmarks, and trace-guided optimization to decide how work should be overlapped and assigned across the iGPU and NPU. The resulting schedule is executed through a unified runtime with shared weights, cross-device synchronization, and NPU-aware queue management. The central challenge is therefore not merely to map compatible operations onto multiple devices, but to expose and schedule heterogeneous opportunity within the compiled C++ execution path under real system constraints.

Our roofline analysis shows why this distinction matters. If heterogeneous execution is restricted only to operations that are directly compatible across the iGPU and NPU, then the achievable speedup is fundamentally bounded by the fraction of runtime spent in those operations. While this limitation is less restrictive at shorter contexts, it becomes increasingly important for longer prompts. For example, in Llama3-8B~\cite{llama}, non-GEMM operations account for only a small fraction of runtime at short prompts, but at a prompt length of 16,384 they can grow to as much as 35\% of total execution time. As a result, heterogeneous GEMM execution alone cannot fully realize the available opportunity, especially across SoCs with widely different iGPU-to-NPU balance points.

HeteroMosaic addresses this in two steps. It first restructures execution to expose additional graph-level heterogeneous opportunity, then applies operation-level partitioning and trace-guided co-optimization to realize that opportunity on the true critical path. Supporting this in practice requires a unified runtime with cross-device and cross-stream synchronization, NPU schedule management, and dynamic iGPU/NPU execution optimizations. HeteroMosaic therefore combines PyTorch C++ tensor and weight management with open-source IRON-based fused NPU kernels and compatible iGPU kernels tailored to the shared heterogeneous execution path.

\begin{figure}[t]
    \centering

    \begin{subfigure}{\columnwidth}
        \centering
        \includegraphics[width=\linewidth]{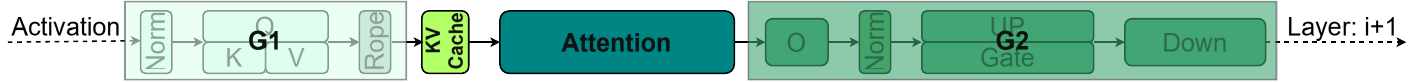}
        \caption{Illustration of our Abstracted LLM Graph.}
        \label{fig:gabstraction}
    \end{subfigure}

    \begin{subfigure}{\columnwidth}
        \centering
        \includegraphics[width=\linewidth]{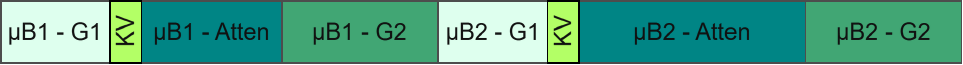}
        \caption{Illustration of Serial Micro-Batching.}
        \label{fig:ubatch}
    \end{subfigure}

    \begin{subfigure}{\columnwidth}
        \centering
        \includegraphics[width=\linewidth]{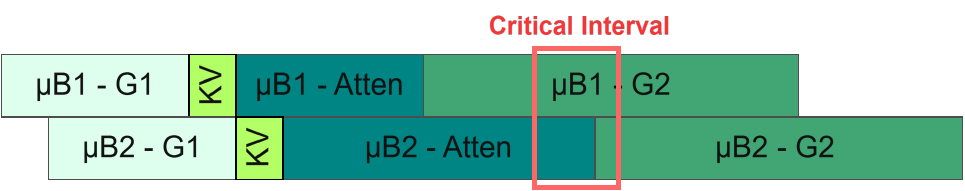}
        \caption{Illustration of Parallel Micro-Batch Dispatch.}
        \label{fig:ubatch_parallel}
    \end{subfigure}

    \begin{subfigure}{\columnwidth}
        \centering
        \includegraphics[width=\linewidth]{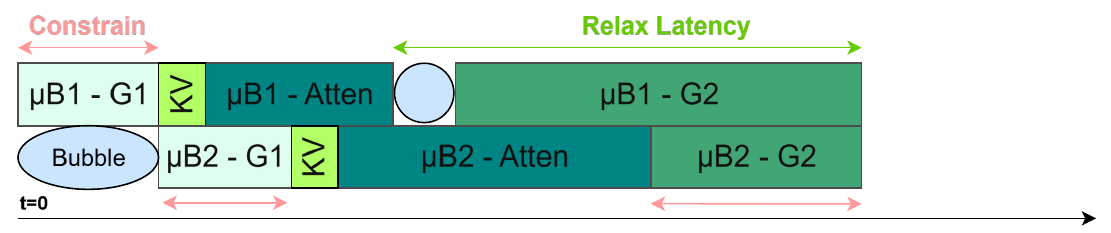}
        \caption{Example of a Latency-Shaped Micro-Batched Schedule.}
        \label{fig:ub_tp_tuned}
    \end{subfigure}

    \caption{Illustration of Heterogeneous Micro-Batching.}
    \label{fig:three_vertical}
\end{figure}

\subsection{Heterogeneous Tensor Parallelism}
\label{sec:hetero_tensorparallel}

Tensor parallelism is the natural first step for heterogeneous execution because it distributes a single operation across multiple compute engines. As introduced in Section~\ref{sec:tensorparallel}, GEMM has three natural split dimensions: $M$-, $K$-, and $N$-splits. On heterogeneous SoCs, these splits can be extended across accelerators when an operation exposes sufficient data parallelism and both backends can execute the same computation. In such cases, a single GEMM can be partitioned across the iGPU and NPU rather than assigned entirely to one device.

Although conceptually simple, heterogeneous splitting remains nontrivial on a unified-memory SoC because unified memory only removes data-copy overhead, not coordination overhead. The accelerators can access the same model buffers without full weight duplication, but the profitability of a split still depends on shared-memory contention, synchronization cost, accelerator balance, backend shape sensitivity, and runtime state. Thus, the best split cannot be inferred from peak TOPS or total arithmetic work alone; it must be characterized empirically and selected in the context of the global schedule.

\subsubsection{Split-dimension trade-offs under unified memory.}
Prior heterogeneous LLM frameworks for edge SoCs use tensor-level partitioning, but do not fully characterize how split dimensions behave under unified-memory contention, DVFS, accelerator imbalance, and backend shape sensitivity~\cite{chen2025heteroinfer}. This matters because $M$-, $K$-, and $N$-splits stress the system differently. For a GEMM $C = AB$, where $A \in \mathbb{R}^{M \times K}$ and $B \in \mathbb{R}^{K \times N}$, an $M$-split partitions rows of $A$ and $C$. This avoids cross-device reduction, but each accelerator must still read the same weight matrix $B$. When $B$ is large and DDR bandwidth is fixed, this duplicated weight traffic can reduce or eliminate the benefit of using both accelerators.

An $N$-split partitions columns of $B$ and $C$. This can avoid duplicate weight reads, but instead duplicates activation reads and can be sensitive to how downstream operators consume the output. If the split introduces extra gathers, transposes, or backend-specific layout conversions, local GEMM speedup may not translate into end-to-end benefits. A $K$-split partitions the reduction dimension, so each accelerator computes a partial result that must later be accumulated. This adds partial-$C$ read/write traffic and a reduction step, but when $K$ is large, the extra merge cost can be amortized by the greater amount of parallel work. Since GEMM kernels are often highly sensitive to the reduction dimension and how it maps onto finite compute resources, $K$-split can become preferable for large-$K$ or large-weight GEMMs despite being more computationally and memory intensive. 

The split choice also introduces accelerator-specific shape sensitivity. A tensor split changes the local GEMM shape seen by each backend, and that shape determines occupancy, tiling efficiency, vector utilization, memory coalescing, and launch overhead. A split that appears balanced by total FLOPs may still leave one backend with a subproblem that is too small, poorly aligned with its tile shape, or unable to saturate its available compute units. Data layout further complicates this trade-off: contiguous DDR access is important for both accelerators, but the iGPU and NPU may prefer different packed layouts, tiling orders, and vectorization patterns.

\subsubsection{Why tensor parallelism alone is insufficient.}
These split-level effects necessitate an empirical characterization of single-operation heterogeneity. A scheduler cannot make accurate global decisions if it assumes that all split-friendly GEMMs behave similarly. It must know which split dimensions are profitable for a given shape, how stable those gains are under DVFS and memory contention, and how much overhead is introduced by synchronization and gather operations. Our microbenchmarks in Section~\ref{sec:microbenchmarks} provide this calibration by measuring $M$-, $K$-, and $N$-split behavior across the three \Ryzen~ balance points.

At the same time, tensor parallelism alone cannot fully realize the heterogeneous opportunity. Operation-level partitioning can improve dense GEMM- and GEMV-like operations either by sharing compute across accelerators or by matching a shape to the device that executes it best. However, this opportunity is not uniformly distributed across the LLM graph. Reduction-heavy, normalization-heavy, and fusion-sensitive operations are weaker targets for operation only sharing; for example, \emph{softmax} introduces row-wise reductions and normalization, while \emph{flash attention}~\cite{dao2022flashattention} relies on fused tiling and on-chip reuse whose benefits can be diluted by partitioning or offload.

As a result, HeteroMosaic treats tensor parallelism as a scheduling primitive rather than a standalone placement rule. The microbenchmarks provide candidate split dimensions and split ratios, but the final choice is made in the context of the global schedule. A split that is fastest for one GEMM in isolation may still hurt end-to-end latency if it increases memory contention, delays a critical stage, or causes downstream layout overhead. HeteroMosaic therefore combines operation-level tensor partitioning with graph-level overlap and critical-path-aware scheduling, rather than greedily optimizing each split-friendly dense layer in isolation.

This limitation motivates HeteroMosaic's graph-level component: rather than treating heterogeneity exclusively as operation-level mapping, we restructure execution to expose new overlap opportunities across the execution graph.

\subsection{Heterogeneity-Aware Scheduling}
Figure~\ref{fig:gabstraction} shows the coarse graph abstraction used by the scheduler. In our implementation, this graph is derived from the compiled PyTorch C++ function-call structure of the model, where nested layer and operator calls are traced into coarse stage nodes such as $G_1$, Attention, and $G_2$. In the original monolithic prefill graph, each layer exposes a mostly serial stage sequence, $G_1 \rightarrow$ Attention $\rightarrow G_2$, over the full prompt. However, this initial graph contains little dependency-valid overlap because there are no independent nodes to schedule. Through micro-batching, we restructure the graph by replicating these nodes per $\mu$B and adding only the KV-cache dependency edges required for causal correctness. As a result, nodes such as $\mu B_1$-$G_2$ and $\mu B_2$-Attention can become independent once the required KV-cache update is complete, exposing overlap that is absent from the original monolithic execution graph.

\subsubsection{Causal Parallel Micro-Batching}

HeteroMosaic uses micro-batching not merely as a memory optimization, but as a mechanism for \emph{exposing concurrent heterogeneous opportunity}. As illustrated in Figure~\ref{fig:gabstraction}, we view each LLM transformer block as three coarse stages: operations before attention that generate and write entries into the KV cache ($G_1$), the attention computation, and operations that follow attention ($G_2$). This abstraction lets us describe prefill not as one monolithic pass over the full prompt, but as a sequence of smaller causally ordered chunks, or micro-batches ($\mu$Bs). Each $\mu$B processes only a subset of prompt tokens, writes its new keys and values into the KV cache, and allows later $\mu$Bs to attend to the tokens that have already been processed. This preserves the same causal semantics as the original prefill, because a later $\mu$B can read earlier KV-cache entries but never depends on future tokens. Figure~\ref{fig:ubatch} shows the conventional serial form used by frameworks such as \texttt{llama.cpp}: $\mu B_1$ executes first and extends the KV cache, then $\mu B_2$ executes using the deeper causal history created by $\mu B_1$, and the final output is produced after the last micro-batch. Later $\mu$Bs therefore attend over a longer KV history and incur longer attention time.

Our first scheduling insight is that these micro-batches can themselves execute in parallel. However, to preserve attention causality, a subsequent $\mu B$ may only read from the KV cache after the preceding $\mu B$ has written its entries. Figure~\ref{fig:ubatch_parallel} illustrates the resulting schedule. Compared with the serial schedule in Figure~\ref{fig:ubatch}, $\mu B_2$ does not need to wait for all of $\mu B_1$ to finish; once $\mu B_1$ has produced the required KV-cache entries, $\mu B_2$ can begin its attention stage while $\mu B_1$ continues into $G_2$. Thus, work such as $\mu B_1$-$G_2$ can overlap with $\mu B_2$-Attention, creating heterogeneous overlap that is absent in the original monolithic schedule.

HeteroMosaic realizes this overlap using a small pool of CPU work-dispatch threads that arbitrate dependency-ready micro-batch stages. Each dispatch thread selects ready work from the global schedule and forwards it to the appropriate backend. iGPU work is issued asynchronously through independent HIP streams, with \texttt{hipEventRecord}~\cite{ROCm_HIP_Event} used to enforce cross-stream dependencies at KV-cache update points. NPU work is not dispatched directly through additional general-purpose work threads; instead, it is enqueued to a dedicated NPU management thread, which serializes and coalesces NPU requests before issuing them to the NPU runtime, as described in Section~\ref{sec:NPUruntime}.

The number of work-dispatch threads is therefore a tunable runtime parameter rather than a fixed design choice. It depends on the available CPU cores, iGPU compute units, and platform power budget. In our evaluated systems, the \Ryzen~7~350 and \Ryzen~9~HX~370 use two work-dispatch threads, while the larger \Ryzen~Max+~395 uses three. The additional CPU resources, larger iGPU, and higher power budget on the \Ryzen~Max+~395 allow more independently dispatched GPU work to improve iGPU occupancy without overwhelming host-device coordination, while the NPU remains managed through its single dedicated queueing path.

Importantly, parallel micro-batching alone does not necessarily reduce end-to-end LLM latency on a single-accelerator system, since all work still contends for the same execution resources. In our experiments, asynchronous parallel dispatch of micro-batches could even hurt performance due to cache-level interference; for example, $\mu B_1$-$G1$ and $\mu B_2$-$G1$ may simultaneously read duplicate QKV weights into the iGPU LLC, leading to cache thrashing. In a heterogeneous system, however, these same causally valid micro-batches can be separated across devices and converted into useful overlap. This connects directly back to the roofline model: the roofline identifies where heterogeneous opportunity exists in principle, while micro-batching reshapes the execution graph to expose schedulable opportunity that is hidden in the original monolithic schedule. In this sense, micro-batching does not change the analytical bound itself; rather, it makes more of the bound reachable in practice by uncovering previously hidden heterogeneous opportunity.

\begin{algorithm}[t]
\caption{Runtime-Aware Latency Shaping}
\label{alg:single_node_latency_shaping}
\centering
\begin{minipage}{0.98\columnwidth}
\begin{algorithmic}[1]
\State \textbf{Input:} node $n$, competing stages $C_w$
\State \textbf{Output:} edit $q_n$

\State $S \gets \Call{GetSystemInfo}{}$
\Comment{DVFS, device balance, NPU runtime}
\State $q_n \gets \Call{NoEdit}{}$
\Comment{default edit}
\State $b_n \gets 0$
\Comment{no bubble by default}

\If{\textbf{not} $\Call{IsHeteroCompatible}{n}$}
    \If{\textbf{not} $\Call{IsCritical}{n}$}
        \State $b_n \gets \Call{ChooseBubble}{n, C_w, S}$
        \State $q_n \gets \Call{InsertBubble}{n, b_n}$
    \EndIf
    \State \Return $q_n$
\EndIf

\State $\beta_n \gets \Call{TargetBalancePoint}{n, S}$
\Statex \Comment{e.g., turbo $\rightarrow 0.5$, sustained $\rightarrow 0.6$}
\State $s_n \gets \Call{QuantizeSplit}{n, \beta_n}$

\If{$\Call{IsCritical}{n}$}
    \State $s_n \gets \Call{BiasTowardMaxResources}{n, s_n, S}$
\Else
    \State $s_n, b_n \gets \Call{BiasBubbleAwayFromCriticalNodes}{n, s_n, S}$
\EndIf

\State $q_n \gets \Call{SetSplitAndBubble}{n, s_n, b_n}$
\State \Return $q_n$
\end{algorithmic}
\end{minipage}
\end{algorithm}

\subsubsection{Heterogeneous Critical-Interval Latency Shaping}
\label{sec:latencyshaping}

Once causal parallel micro-batching has exposed additional graph-level overlap, HeteroMosaic uses heterogeneous allocation as a \emph{latency-shaping} mechanism that controls when a \emph{node} in the task graph completes relative to the global schedule. The key control is the allocation choice for that node. A tensor-parallel split close to the device-optimal balance point can decrease the node's execution time by using both accelerators effectively. For non-critical nodes, however, HeteroMosaic can choose slower but less interfering allocations: single-accelerator execution on the fastest accelerator, single-accelerator execution on the second-fastest accelerator, or a deliberately less aggressive split. Thus, the allocation space spans fast tensor-parallel configurations for critical nodes and slower single-device or weakly split configurations for non-critical nodes. Algorithm~\ref{alg:single_node_latency_shaping} formalizes this policy for a single node within a \emph{critical interval}, a window in time in which dependency-ready stages from overlapping micro-batches execute concurrently and contend for the same SoC resources. Within such an interval, HeteroMosaic contracts nodes on the critical path while relaxing non-critical nodes through weaker allocations or bubbles, freeing resources for the stages that determine interval completion. Figure~\ref{fig:ubatch_parallel} shows the overlapping micro-batch structure that creates these intervals, and Figure~\ref{fig:ub_tp_tuned} illustrates how latency shaping changes stage completion times within them.

The key idea in Algorithm~\ref{alg:single_node_latency_shaping} is that heterogeneous allocation is both \emph{runtime-aware} and dependency-preserving. Candidate nodes are considered in topological order over the traced micro-batched DAG, while the competing set $C_w$ captures the stages that overlap with the current node inside the critical interval. This topological order preserves producer--consumer constraints, such as KV-cache writes before later attention reads; it does not serialize execution, since dependency-ready nodes from different micro-batches may still execute concurrently. Figure~\ref{fig:ub_tp_tuned} illustrates how the policy uses this dependency-valid overlap. In the constrained region, critical stages such as $\mu B_1$-$G_1$ are biased toward faster allocations so that they complete earlier and unblock dependent work. In the relaxed-latency region, non-critical stages such as $\mu B_2$-$G_2$ may use a less aggressive allocation or an inserted bubble so that they finish just in time without extending the global critical path.

Given a node and its competing stages, the algorithm first checks whether the node is heterogeneity-compatible. If not, the only remaining control is whether to insert a bubble for a non-critical node so that it interferes less with the critical path, as shown by the delayed $\mu B_2$-$G_1$ stage in Figure~\ref{fig:ub_tp_tuned}. If the node is compatible, the runtime queries the current system state and selects a target balance point before quantizing that decision into a concrete split. This system state includes DVFS behavior, effective iGPU--NPU asymmetry, and NPU runtime effects such as queueing and configuration pressure.

The concrete latency-shaping decision can therefore change across SoC variations and runtime regimes. In particular, the balance point can shift between a short-lived \emph{turbo} regime, in which the SoC operates at temporarily elevated power and frequency, and a \emph{sustained} regime, in which the chip settles to a lower long-term operating point under thermal and power constraints. A split that is favorable while the SoC remains in turbo may no longer be optimal once execution enters the sustained-power regime, and NPU-side effects such as reconfiguration pressure can further bias the preferred allocation. HeteroMosaic therefore does not use a fixed split rule; instead, as shown in Algorithm~\ref{alg:single_node_latency_shaping}, it selects a runtime-aware target balance point and then biases the final split differently depending on whether the node is critical or non-critical.

This distinction matters because graph-level overlap becomes substantially more valuable when combined with operation-level latency shaping. By adjusting the allocation of work across the iGPU and NPU, HeteroMosaic can control the execution time of stages such as $G_1$ and $G_2$ so that they better align with the global graph schedule. Critical stages are biased toward allocations that minimize completion time under the current system state, while non-critical stages may use less aggressive splits, single-accelerator execution, or bubbles so that they do not interfere with more important work. In this way, heterogeneous allocation is used not merely to distribute computation, but to shape node latency around the critical path under practical runtime effects.

\begin{figure}[t]
    \centering

    \begin{subfigure}{\columnwidth}
        \centering
        \includegraphics[width=\linewidth]{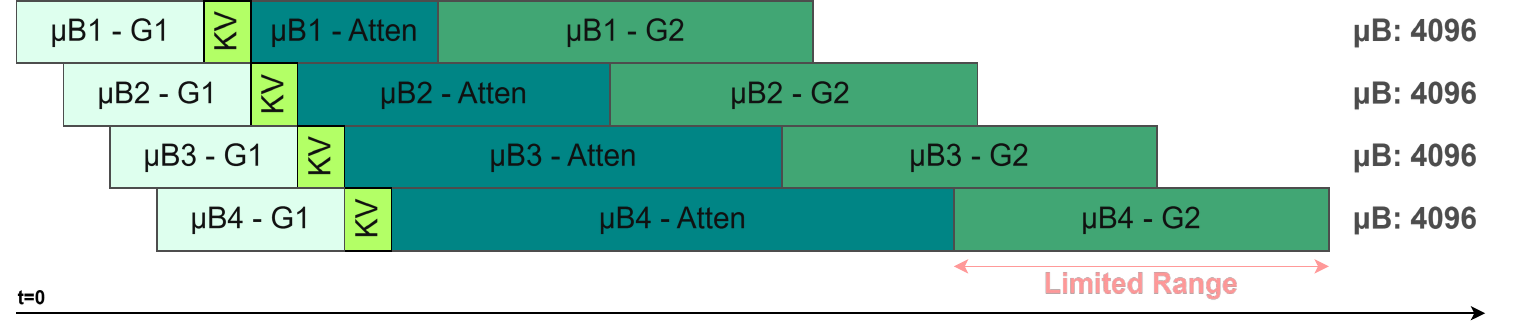}
        \caption{Limits of Micro-Batching and Latency Shaping.}
        \label{fig:ub_deep}
    \end{subfigure}

    \begin{subfigure}{\columnwidth}
        \centering
        \includegraphics[width=\linewidth]{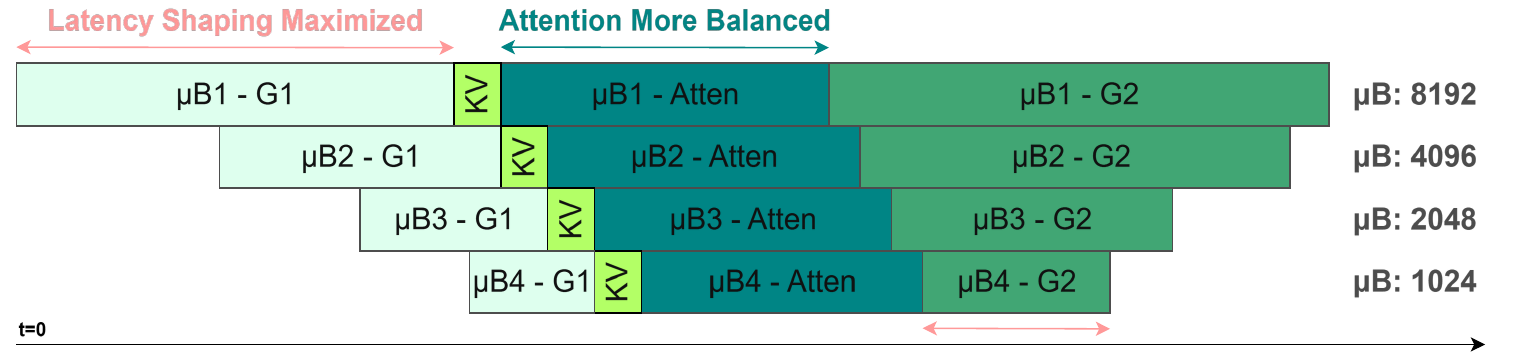}
        \caption{Example of an Asymmetric Schedule.}
        \label{fig:ub_deep_chunked}
    \end{subfigure}

    \caption{Example of Asymmetric Micro-Batching Schedules.}
    \label{fig:two_vertical}
\end{figure}

\subsubsection{Asymmetric Micro-Batching for Long Contexts}

The fixed micro-batch schedules discussed so far implicitly assume that heterogeneity provides sufficient dynamic range to align node completion times with the global schedule. This assumption becomes increasingly fragile at long context lengths and SoCs with accelerator asymmetry. In Transformer-based graphs, the root cause is the different scaling behavior of linear projections and attention. For a fixed model width, the computational complexity of linear projections scales linearly with prompt length, while the computational complexity of attention scales quadratically. More concretely, projection layers scale as $O(Nd^2)$, whereas attention scales as $O(N^2d)$, where $N$ is the prompt length and $d$ is the embedding dimension. As a result, when $N$ is small, non-attention computation occupies a larger share of the schedule, but as $N$ grows, attention increasingly dominates. As illustrated in Figure~\ref{fig:ub_deep}, exaggerated here for clarity, naive micro-batching of long prompts can produce schedules in which the growing attention time, driven both by the intrinsic complexity of attention and by the deeper KV history seen by later micro-batches, overwhelms the latency-shaping range available through heterogeneous tensor parallelism. This challenge becomes even more severe on unbalanced systems, where the achievable heterogeneous dynamic range may be insufficient to satisfy the schedule induced by the graph.

The limitation is that, for long prompts, a fixed micro-batch schedule can create stage imbalance that heterogeneous latency shaping cannot fully absorb. As later micro-batches see a deeper KV history, their attention stages grow and begin to dominate the micro-batched schedule. The maximum acceleration each micro-batch can achieve is bounded by the \emph{dynamic range} of the latency-shaping mechanism. Here, dynamic range refers to how much the runtime can change a node's execution time through tensor-parallel splits, single-accelerator assignment, or bubbles. This range is larger on balanced systems: if the iGPU and NPU have comparable capability, using both accelerators can ideally provide close to a $2\times$ latency reduction over either single-accelerator baseline. On a skewed system, however, the range relative to the stronger accelerator is much smaller. For example, with a $1{:}4$ NPU-to-iGPU compute ratio, the ideal heterogeneous speedup over the iGPU is only $(1+4)/4=1.25\times$. Thus, when attention growth exceeds this limited shaping range, the runtime cannot contract the long-attention micro-batches enough to achieve the same degree of speedup observed at shorter prompt lengths.

HeteroMosaic addresses this limitation by expanding the search space to include not only heterogeneous allocation, but also micro-batch size selection. Rather than assuming a fixed, equal-size micro-batch schedule, it identifies a micro-batch strategy that better matches the available latency-shaping range. The intuition is illustrated in Figure~\ref{fig:ub_deep_chunked}, which replaces the fixed schedule in Figure~\ref{fig:ub_deep} with an example gradually decreasing micro-batch size. In the fixed schedule, later micro-batches spend disproportionate time in attention because attention cost grows with the deeper KV history. The decreasing schedule counteracts this by making later micro-batches smaller, reducing the attention time that would otherwise dominate those stages. Larger earlier micro-batches preserve useful projection work when the attention history is still shallow. This better balances attention time across micro-batches and reduces the amount of linear projection work that remains on the critical path for deeper micro-batches. Although this strategy is not always the winning solution, we find it particularly useful on systems with limited heterogeneous range and for long prompt lengths, where fixed schedules can otherwise become too rigid.

\subsubsection{Critical-Interval Co-Optimization}
\label{sec:criticalintercalco}

The exposed overlap creates a \emph{non-greedy} schedule optimization problem: locally faster mappings can worsen end-to-end latency by perturbing more critical stages. In a heterogeneous micro-batched schedule, graph-level overlap and operation-level allocation are inherently coupled. For example, in Figure~\ref{fig:ubatch_parallel}, greedily optimizing $\mu B_2$-$G_2$ can negatively impact $\mu B_1$-$G_1$, and vice versa, because both execute within a shared runtime environment and contend for the same heterogeneous resources. As a result, local improvements do not necessarily translate into global gains. HeteroMosaic therefore introduces a trace-guided critical-interval co-optimization algorithm that leverages its explicit C++ runtime to emit fine-grained execution traces. Each trace record contains the stage identifier, micro-batch identifier, backend assignment, tensor split ratio, start timestamp, end timestamp, and measured latency. These records are then used to evaluate graph-level overlap and operation-level resource allocation against measured end-to-end latency.

Algorithm~\ref{alg:trace_guided_cosearch} summarizes this procedure. We adopt a trace-driven approach because graph nodes in real systems can skew in time due to scheduler race conditions, DVFS, and runtime-specific behaviors. Since the algorithm depends on accurately identifying the true critical interval, relying on a static schedule or purely analytical timing model proved insufficient. As shown in lines~3--8 of Algorithm~\ref{alg:trace_guided_cosearch}, the method first evaluates candidate micro-batch schedules, executes each one, and constructs an internal DAG from the resulting runtime trace. It then selects the schedule that minimizes measured critical-path pressure.

After this initialization step, the algorithm repeatedly builds \emph{critical intervals}, i.e., time windows in which the same set of stage instances are simultaneously active and therefore contend for the same GPU/NPU environment. Lines~11--17 of Algorithm~\ref{alg:trace_guided_cosearch} then iterate over nodes in the current traced DAG and invoke Algorithm~\ref{alg:single_node_latency_shaping} to propose a node-level edit. Each edit applies the \emph{latency-shaping} mechanism described in Section~\ref{sec:latencyshaping} in the context of the measured global schedule, adjusting a node's split, accelerator assignment, or bubble so that the node completes earlier or later as needed to reduce critical-path pressure rather than optimizing the node in isolation. In this way, Algorithm~\ref{alg:single_node_latency_shaping} provides the local runtime-aware policy, while Algorithm~\ref{alg:trace_guided_cosearch} decides whether that local edit is globally beneficial.

Within each interval, HeteroMosaic jointly optimizes two decisions: the heterogeneous allocation, and optional bubbles for non-critical nodes, which intentionally delay execution when doing so protects the global critical path. Every proposed edit is re-executed and accepted only if it improves measured end-to-end latency, as shown in lines~13--16 of Algorithm~\ref{alg:trace_guided_cosearch}. This structure is important because a node-level improvement from Algorithm~\ref{alg:single_node_latency_shaping} does not necessarily reduce overall latency once the full schedule is re-executed. This tuning is performed offline once per model/device configuration. In our current implementation, the full trace-guided search takes on the order of eight hours and produces reusable configurations for representative prompt-length ranges. At deployment time, HeteroMosaic selects the corresponding configuration and does not repeat the full search. Thus, the search cost is paid during offline calibration rather than on the latency-critical inference path.

\begin{algorithm}[t]
\caption{Trace-Guided Critical-Interval Co-Optimization}
\label{alg:trace_guided_cosearch}
\centering
\begin{minipage}{0.98\columnwidth}
\begin{algorithmic}[1]
\State \textbf{Input:} initial configuration $C_0$, micro-batch schedule $S_0$
\State \textbf{Output:} tuned configuration $(C^\star, S^\star)$

\ForAll{$s \in \mathcal{S}_0$}
\State $(T_s, L_s) \gets \Call{RunModelAndCollectTrace}{C_0, s}$
\Statex \hspace{\algorithmicindent} $\triangleright$ Execute schedule $s$ and record trace $T_s$ and latency $L_s$

\State $G_s \gets \Call{BuildInternalDAG}{T_s}$
\Statex \hspace{\algorithmicindent} $\triangleright$ Build a stage DAG from traced dependencies and timing

\State $P_s \gets \Call{MeasureCriticalPathPressure}{G_s}$
\Statex \hspace{\algorithmicindent} $\triangleright$ Critical-path pressure metric
    \If{$P_s < P^\star$} \Comment{$P^\star$ initialized to $\infty$}
        \State $S^\star, C^\star, T^\star, G^\star, P^\star, L^\star \gets s, C_0, T_s, G_s, P_s, L_s$
    \EndIf
\EndFor

\For{$i = 1$ to $B$} \Comment{Search budget $B$}
    \State $W \gets \Call{BuildCriticalIntervals}{G^\star}$

    \ForAll{$n \in G^\star$}
        \State $q_n \gets \Call{LatencyShaping}{n, W}$
        \State $C_q \gets \Call{ApplyEdit}{C^\star, q_n}$
        \State $(T_q, L_q) \gets \Call{RunModelAndCollectTrace}{C_q, S^\star}$
        \State $G_q \gets \Call{BuildInternalDAG}{T_q}$

        \If{$L_q < L^\star$}
            \State $C^\star, T^\star, G^\star, L^\star \gets C_q, T_q, G_q, L_q$
        \EndIf
    \EndFor
\EndFor

\State \Return $(C^\star, S^\star)$
\end{algorithmic}
\end{minipage}
\end{algorithm}

\subsection{\Ryzen~ Runtime Realization}
The previous subsection described the scheduling concepts underlying HeteroMosaic. We now describe the \Ryzen-specific runtime mechanisms required to realize them efficiently in practice. These mechanisms are not the primary algorithmic contribution of the paper; rather, they are the platform-specific enablers that allow the scheduling framework to execute effectively on unified-memory \Ryzen~ SoCs.

\subsubsection{Unified Memory and Synchronization}
\label{sec:syncronization}

A primary challenge in mapping compute across the iGPU and NPU on \Ryzen~ is reconciling separate memory and runtime models. On \Ryzen~ running Linux, we implement unified memory across different runtimes through the Direct Rendering Manager (DRM)~\cite{drm} stack, which allows us to create shareable buffers across devices. A second challenge is synchronization across heterogeneous runtimes. On the iGPU, execution follows an asynchronous stream-based model through ROCm~\cite{rocm}, where the host dispatches work non-blockingly and dependencies are preserved within each stream. The NPU, by contrast, follows a host-dispatch-and-return model, where the host submits work and blocks until completion. HeteroMosaic reconciles these differing runtimes by using independent HIP streams to preserve correctness within a micro-batch, while combining \texttt{hipEventRecord} and \texttt{hipStreamWaitValue32}~\cite{ROCm_HIP_Event, ROCm_HIP_StreamWait} as host-mediated signaling mechanisms to preserve dependencies across multiple iGPU streams and the NPU.

\subsubsection{Runtime NPU Management}
\label{sec:NPUruntime}

\Ryzen~ NPUs, when implemented well, can exhibit lower GEMM shape sensitivity than prior mobile NPU paths such as HeteroInfer, where fixed systolic-array tiling and tensor order/shape alignment introduce stage-, order-, and shape-sensitive performance variation~\cite{chen2025heteroinfer}. However, this comes at the cost of fixed reconfiguration overhead. Prior work~\cite{andre-paper} has shown that NPU reconfiguration can negatively impact end-to-end performance. On \Ryzen, the NPU exposes two levels of configuration: (i) \emph{datapath} reconfiguration, which is expensive because it reprograms tile internals and switchbox routes, and (ii) \emph{DMA} reconfiguration, which updates the DMA engines and is effectively negligible even for microsecond-scale kernels. The central challenge is therefore to choose an NPU schedule that maximizes reuse of datapath configurations. This issue becomes even more important in parallel micro-batched schedules, where, without careful alignment, the NPU may thrash between configurations across adjacent micro-batches.

To address this, we implement a dedicated NPU management thread that services requests generated by the main micro-batch schedule shown in Figure~\ref{fig:hetero-flow}. Independent micro-batches enqueue NPU operations into this thread, which then coalesces requests with compatible datapath configurations. In addition, the same operation often appears across multiple micro-batches and differs only in its activation inputs. Without optimization, operations such as $\mu B_1$-$G1$ and $\mu B_2$-$G1$ would be dispatched separately, causing the NPU to reread the same $Q$ weights multiple times. When such operations recur within a suitable interval, the NPU management thread can coalesce these otherwise independent but structurally identical operations, thereby reducing both NPU dispatch overhead and redundant weight reads.

\subsubsection{Custom Heterogeneous GPU Kernels}

The IRON-based NPU kernels available through the open-source fork in~\cite{glassescrab_mlir_aie_2026} rely on a custom weight layout. As a result, to share weights effectively between the iGPU and NPU, HeteroMosaic cannot rely on existing GPU kernels. We therefore develop optimized iGPU kernels whose weight layout is compatible with the NPU layout, eliminating weight duplication across accelerators. Beyond GEMM, we also implement framework-tailored iGPU kernels for non-GEMM stages such as normalization, KV-cache handling, and FlashAttention-style attention~\cite{dao2022flashattention}. In our experiments, these optimized iGPU kernels outperform available W4A16 implementations such as \texttt{llama.cpp}~\cite{llama.cpp}, largely because they are specialized for RDNA-based iGPUs~\cite{amd_wmma_rdna3_2023} and HeteroMosaic's execution path.

\section{Evaluation}

\begin{figure*}[t]
    \centering
    \setlength{\tabcolsep}{0pt}
    \renewcommand{\arraystretch}{1.25}

    \newlength{\plotwMicro}
    \setlength{\plotwMicro}{0.32\textwidth}

    \newcommand{\splitlabel}[1]{%
        \raisebox{-0.05\height}{%
            \rotatebox[origin=c]{90}{\textbf{#1}}%
        }%
    }

    \newcommand{\microplotcell}[1]{%
        \raisebox{-0.5\height}{%
            \includegraphics[width=\plotwMicro,trim=12 4 12 4,clip]{#1}
        }%
    }

    \begin{tabular}{@{} c @{\hspace{4pt}} c @{\hspace{2pt}} c @{\hspace{2pt}} c @{}}
        &
        \multicolumn{1}{c}{\textbf{\Ryzen~ 7 350}} &
        \multicolumn{1}{c}{\textbf{\Ryzen~ 9 HX 370}} &
        \multicolumn{1}{c}{\textbf{\Ryzen~ Max+ 395}} \\

        \splitlabel{M-split} &
        \microplotcell{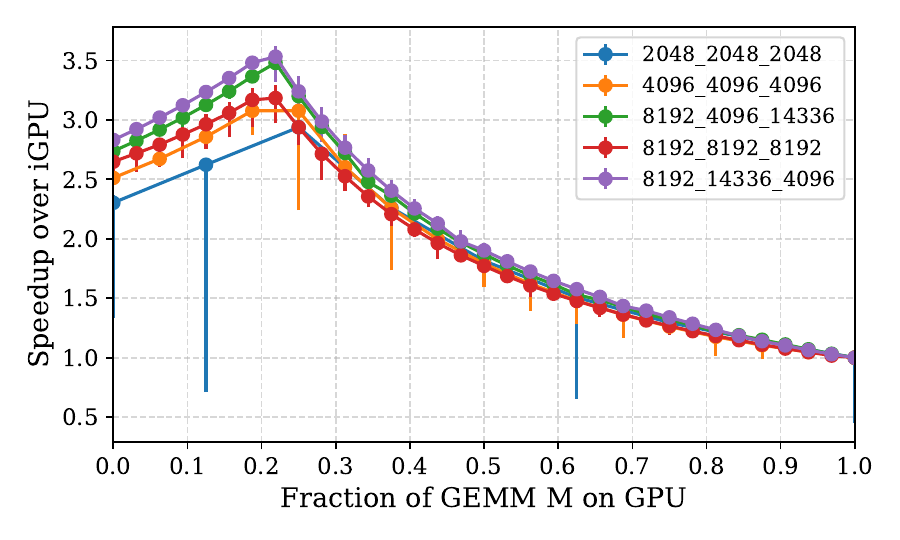} &
        \microplotcell{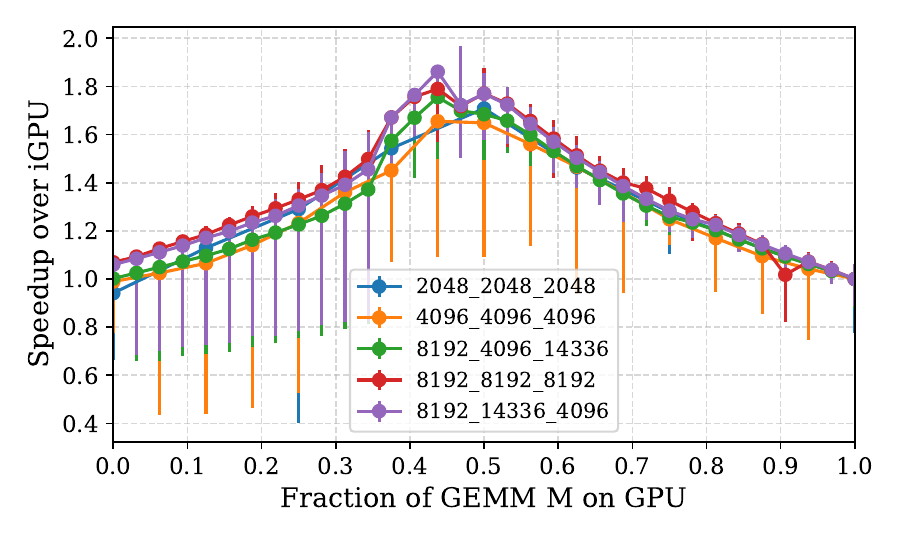} &
        \microplotcell{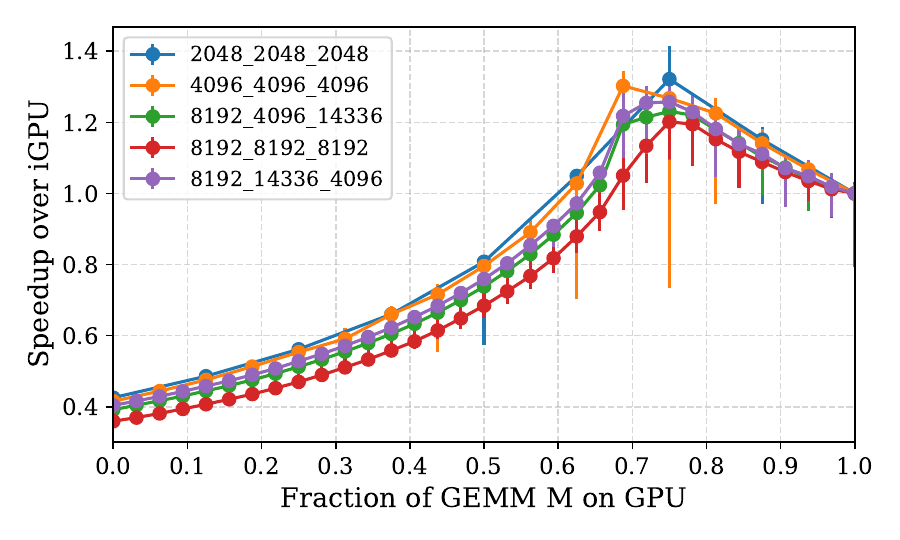} \\

        \splitlabel{K-split} &
        \microplotcell{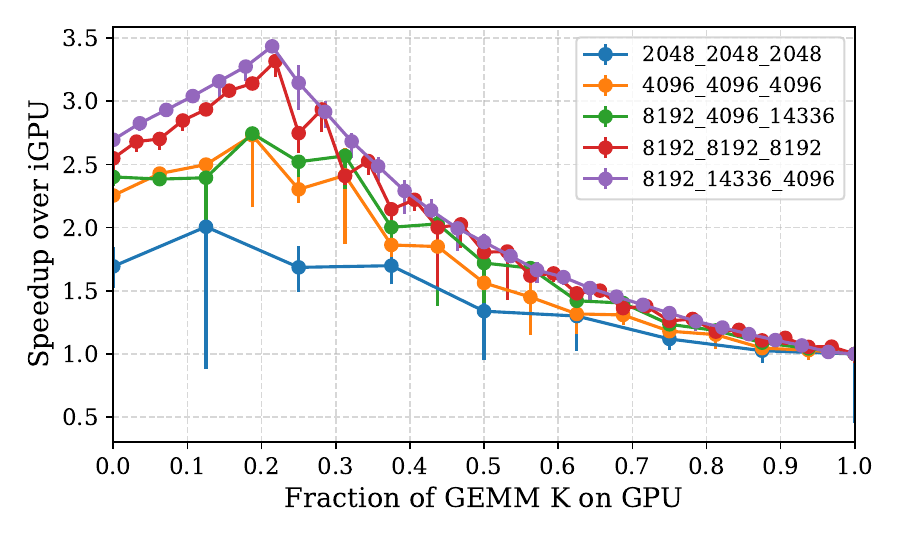} &
        \microplotcell{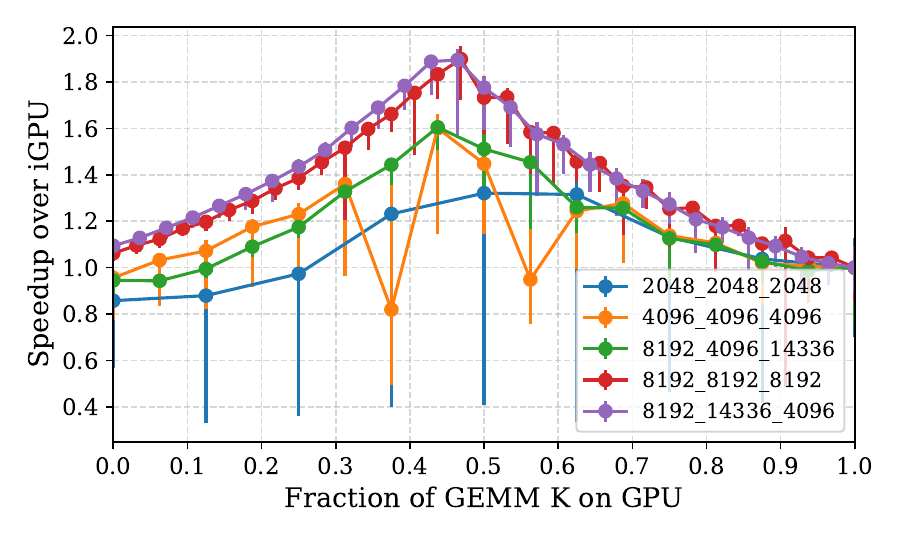} &
        \microplotcell{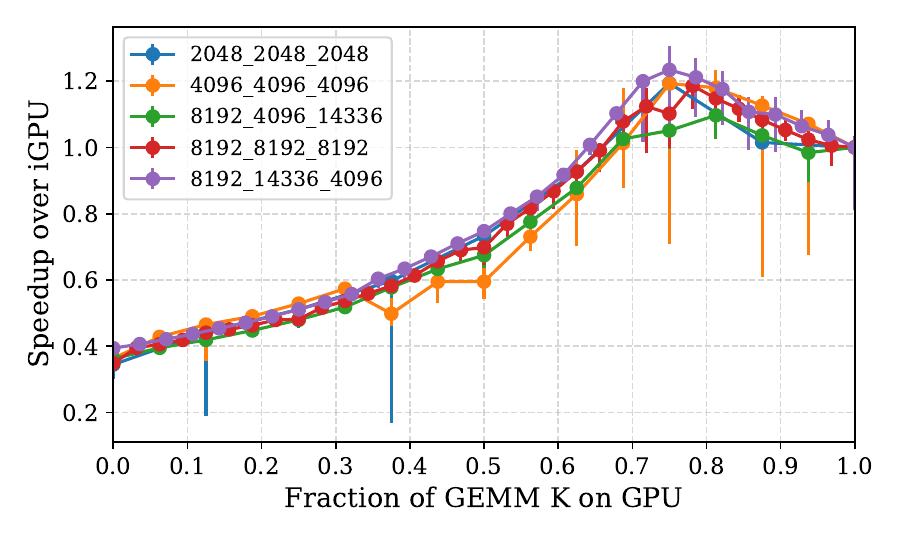} \\

        \splitlabel{N-split} &
        \microplotcell{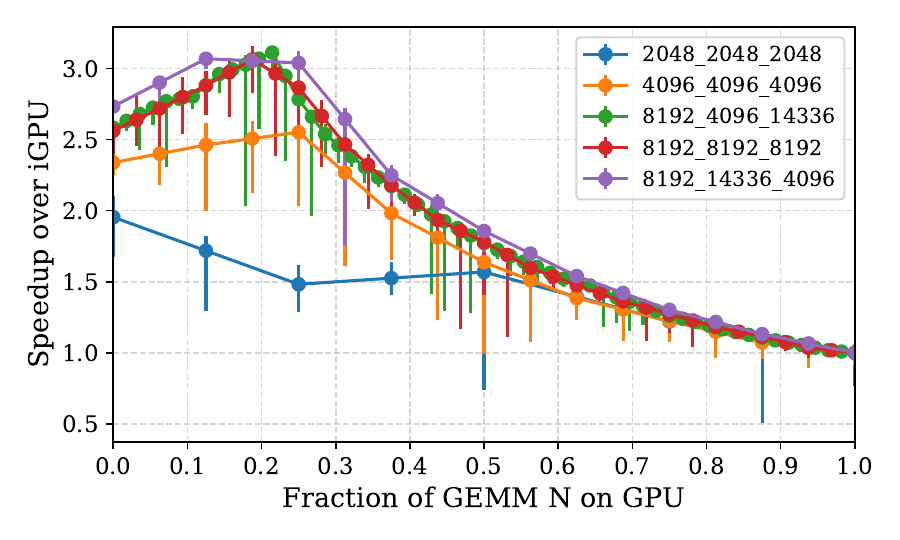} &
        \microplotcell{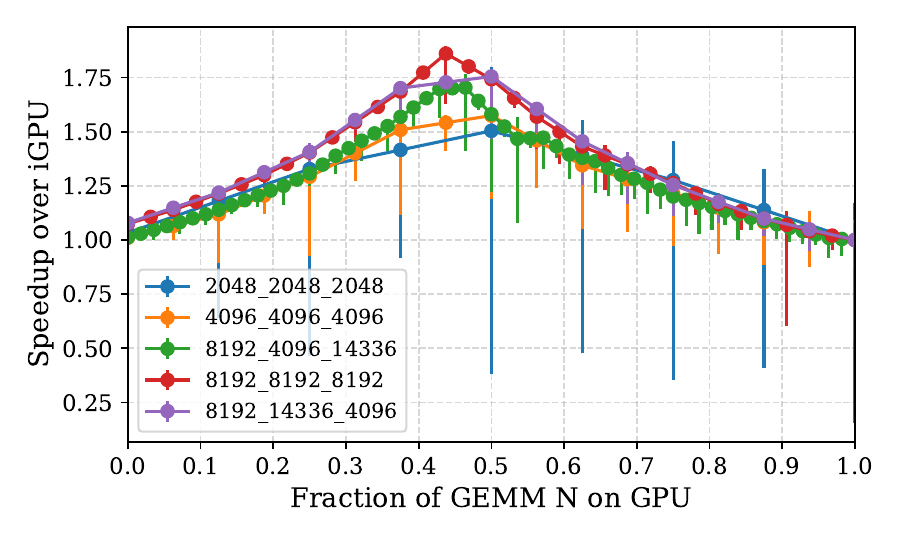} &
        \microplotcell{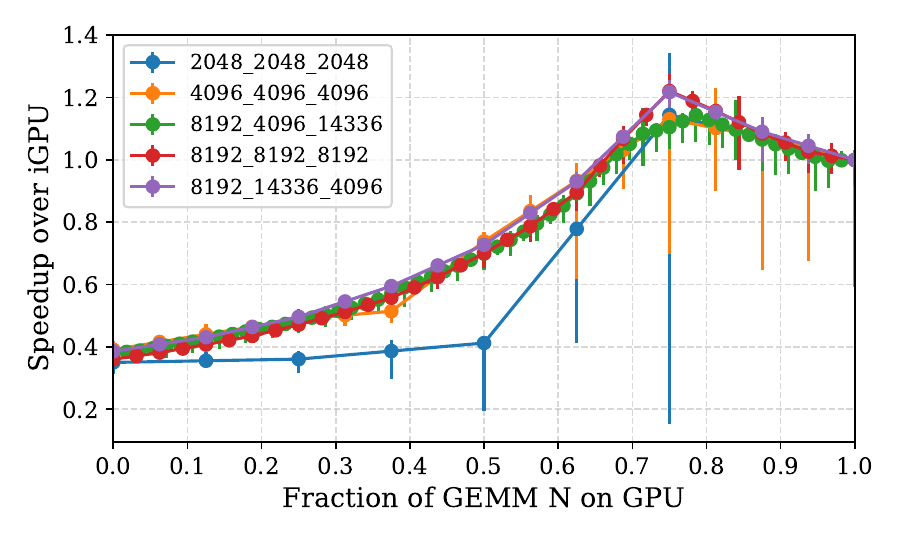} \\
    \end{tabular}
    \caption{Speedup of heterogeneous execution for GEMMs of varying sizes on the NPU and iGPU, normalized to the iGPU baseline. Results are shown for M-, K-, and N-dimension tensor splits across three \Ryzen~systems.}
    \label{fig:microbenchmark}
\end{figure*}

To evaluate HeteroMosaic, we study three current-generation \Ryzen~ devices spanning NPU-heavy, balanced, and iGPU-heavy designs: the \Ryzen~ 7 350, \Ryzen~ 9 HX 370, and \Ryzen~ Max+ 395, respectively. Our evaluation is organized to test the two-stage argument of this paper. First, we use microbenchmarks to measure the directly accelerable core of inference and to evaluate whether the achievable behavior of GEMM aligns with the roofline-based opportunity identified in Section~\ref{sec:roofline}. Second, we use end-to-end LLM inference to test whether HeteroMosaic can expose and recover that opportunity at the graph level under realistic runtime constraints. We then contextualize these results against relevant prior heterogeneous frameworks, analyze the contribution of HeteroMosaic’s individual components through ablations, and study whether the resulting speedups translate into lower energy under the TDP-constrained behavior of modern SoCs.

\subsection{Microbenchmarks}
\label{sec:microbenchmarks}

Microbenchmarks isolate the directly accelerable GEMM core of inference and provide the cleanest measured view of the roofline-identified opportunity. They also quantify the split-dimension design space introduced in Section~\ref{sec:tensorparallel} and Section~\ref{sec:hetero_tensorparallel} by answering four questions. First, do the measured gains match the opportunity predicted by the roofline model? Second, how do those gains vary across SoCs with different iGPU--NPU balance points? Third, which tensor-parallel split dimension is most effective for different GEMM shapes? Fourth, how much of the observed behavior reflects genuine heterogeneous opportunity rather than shape sensitivity, memory contention, or kernel artifacts~\cite{streamk, nvidia_dl_performance_matrix}?

We map GEMM operations heterogeneously using the $M$-, $K$-, and $N$-split strategies described in Section~\ref{sec:tensorparallel}. Figure~\ref{fig:microbenchmark} summarizes the results. Each point is averaged over 1024 GEMM runs, with error bars indicating variation across those repeated runs. This variation primarily reflects SoC power-management behavior, where execution can move between a transient turbo/high-power state and a sustained TDP-limited state. Shorter GEMMs are especially sensitive to which regime they execute under, which is evident across platforms for shapes such as $2048{\times}2048{\times}2048$. Longer GEMMs can also show variation when a run spans a turbo-to-sustained transition or begins from a different thermal/power state, as seen for larger shapes such as $8192{\times}14336{\times}4096$ on the \Ryzen~9 HX 370. Overall, the measured trends follow the roofline intuition: the NPU-heavy \Ryzen~7 350 exposes the largest heterogeneous speedups over the iGPU, while the balanced \Ryzen~9 HX 370 shows moderate but still substantial gains, and the iGPU-heavy \Ryzen~Max+ 395 shows the smallest headroom because the iGPU baseline is already strong.

Across the full sweep, $M$-split is the most consistently effective strategy. $M$-split avoids cross-device reduction and works well for small and moderate GEMMs where duplicated weight reads do not dominate execution. However, on the \Ryzen~7 350 and \Ryzen~9 HX 370, where DDR bandwidth and iGPU compute capability are more limited than on the \Ryzen~Max+ 395, $K$-split can outperform $M$-split for larger GEMMs. Using the notation $M{\times}K{\times}N$ for GEMM shapes, with $A \in \mathbb{R}^{M \times K}$, $B \in \mathbb{R}^{K \times N}$, and $C \in \mathbb{R}^{M \times N}$ for $C=AB$, examples of large GEMMs where $K$-split becomes favorable include $8192{\times}8192{\times}8192$ and $8192{\times}14336{\times}8192$. In these cases, splitting along $K$ creates subproblems that better utilize both accelerators, despite the additional reduction step.

The $K$-split results also illustrate why empirical calibration is necessary. Several GEMM shapes, including $2048{\times}2048{\times}2048$, $4096{\times}\allowbreak4096{\times}4096$, and $8192{\times}4096{\times}14336$, perform noticeably worse than what a simple FLOP-balanced split would suggest. This behavior is consistent with backend shape sensitivity: the split creates local GEMM shapes that map awkwardly onto the backend tiling and compute resources, reducing occupancy or tiling efficiency even when the total arithmetic work appears balanced. This effect is most pronounced on the \Ryzen~9 HX 370 for the $4096{\times}4096{\times}4096$ GEMM, which shows sharp performance drops at certain split ratios, such as around 37\% and 56\% GPU assignment. Thus, the best split cannot be inferred from FLOP balance alone; it depends on how the partitioned shape maps onto the physical compute resources of each backend.

$N$-split is generally the weakest strategy in our measurements. Although it can reduce duplicate weight reads in principle, it interacts poorly with the packed weight layout used by our heterogeneous kernels. Because $N$-split partitions columns of $B$ and $C$, it can require strided weight reads and scatter-like writes into the final $C$ tensor. These non-contiguous accesses reduce memory coalescing and introduce gather/scatter overhead, often preventing $N$-split from converting its reduced weight traffic into measured speedup. On the \Ryzen~Max+ 395, $M$-split remains best across the evaluated shapes because higher memory bandwidth and a stronger iGPU make simple output-row partitioning more effective than the reduction overhead of $K$-split or the gather/scatter overhead of $N$-split.

These results provide the empirical bridge between the roofline model and the HeteroMosaic scheduler. For compute-heavy GEMMs, the measured iGPU--NPU speedups track the roofline model's predicted opportunity, while deviations mainly reflect shape sensitivity, memory contention, and runtime overheads. However, the best tensor-parallel strategy still depends on GEMM shape, split ratio, SoC balance, memory bandwidth, layout compatibility, and accelerator-specific shape sensitivity. HeteroMosaic therefore uses these microbenchmarks to calibrate candidate splits, while the global scheduler ultimately determines which split that improves the end-to-end critical path.

\subsection{End-to-End LLM Evaluation}

\begin{figure*}[t]
    \centering
    \setlength{\tabcolsep}{0pt}
    \renewcommand{\arraystretch}{1.5} 

    \newlength{\plotw}
    \setlength{\plotw}{0.32\textwidth} 

    \newcommand{\modellabel}[1]{%
        \raisebox{-0.05\height}{%
            \rotatebox[origin=c]{90}{\textbf{#1}}%
        }%
    }
    
    \newcommand{\plotcell}[1]{%
        \raisebox{-0.5\height}{%
            \includegraphics[width=\plotw,trim=4 4 4 4,clip]{#1}%
        }%
    }
    
    \begin{tabular}{@{} c @{\hspace{4pt}} c @{\hspace{2pt}} c @{\hspace{2pt}} c @{}}
        &
        \multicolumn{1}{c}{\textbf{\Ryzen~ 7 350}} &
        \multicolumn{1}{c}{\textbf{\Ryzen~ 9 HX 370}} &
        \multicolumn{1}{c}{\textbf{\Ryzen~ Max+ 395}} \\

        \modellabel{Phi3.5-3.8B} &
        \plotcell{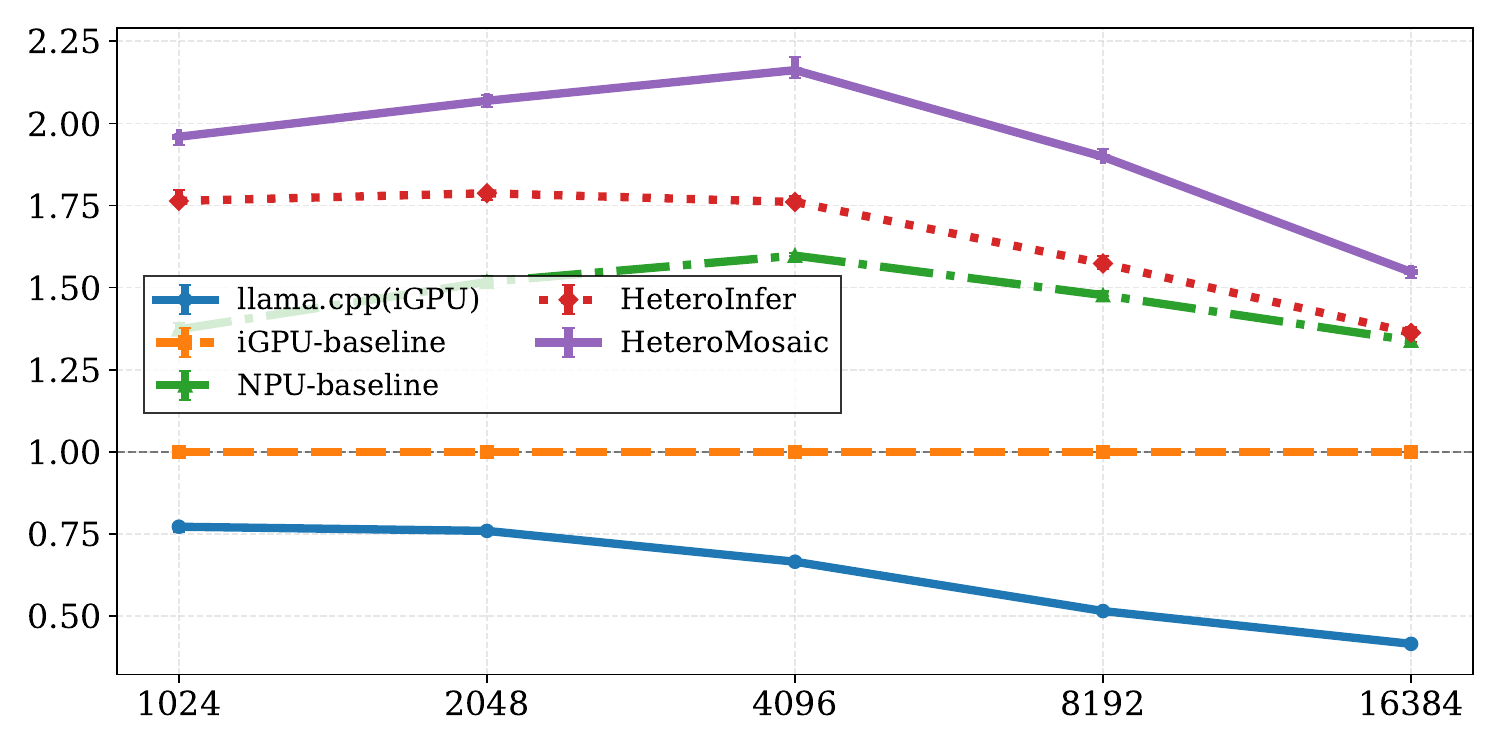} &
        \plotcell{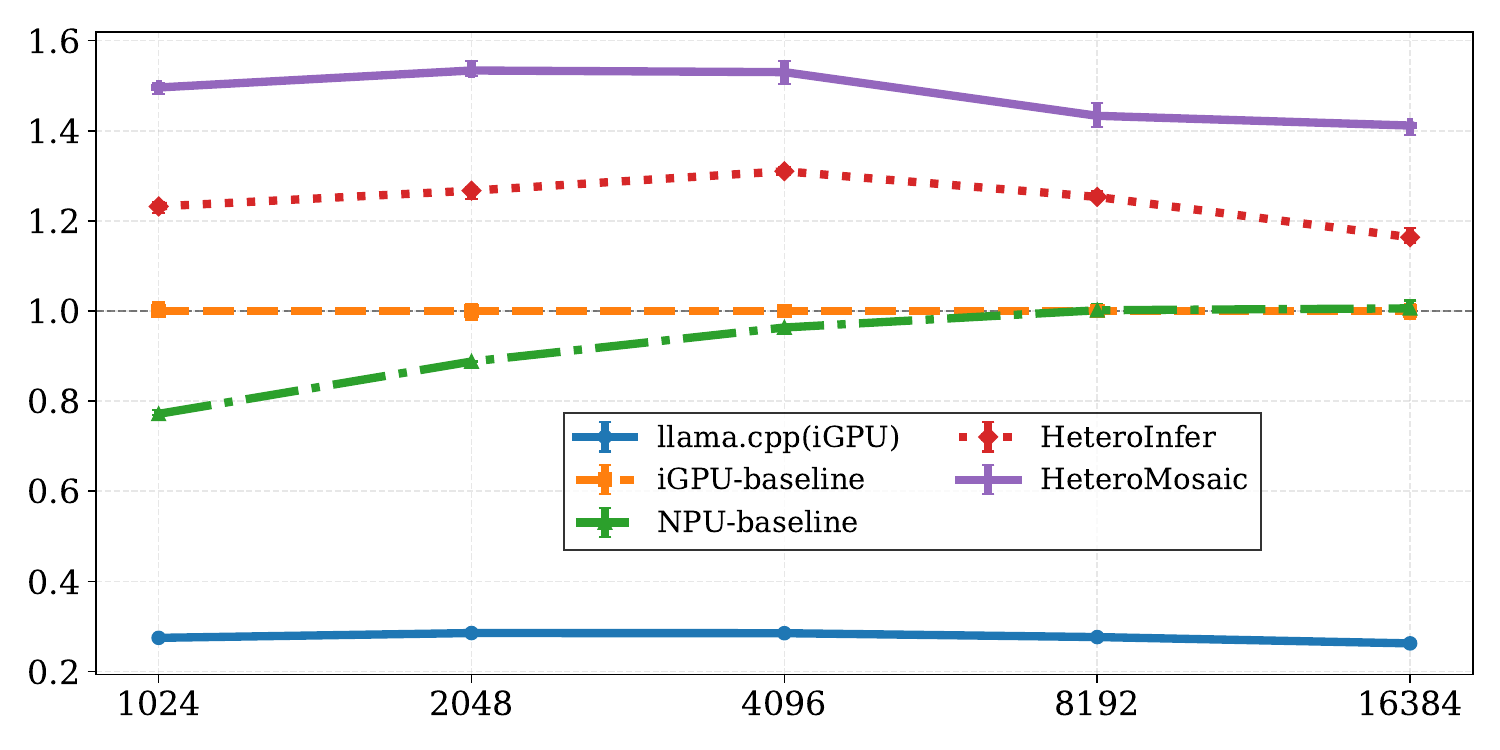} &
        \plotcell{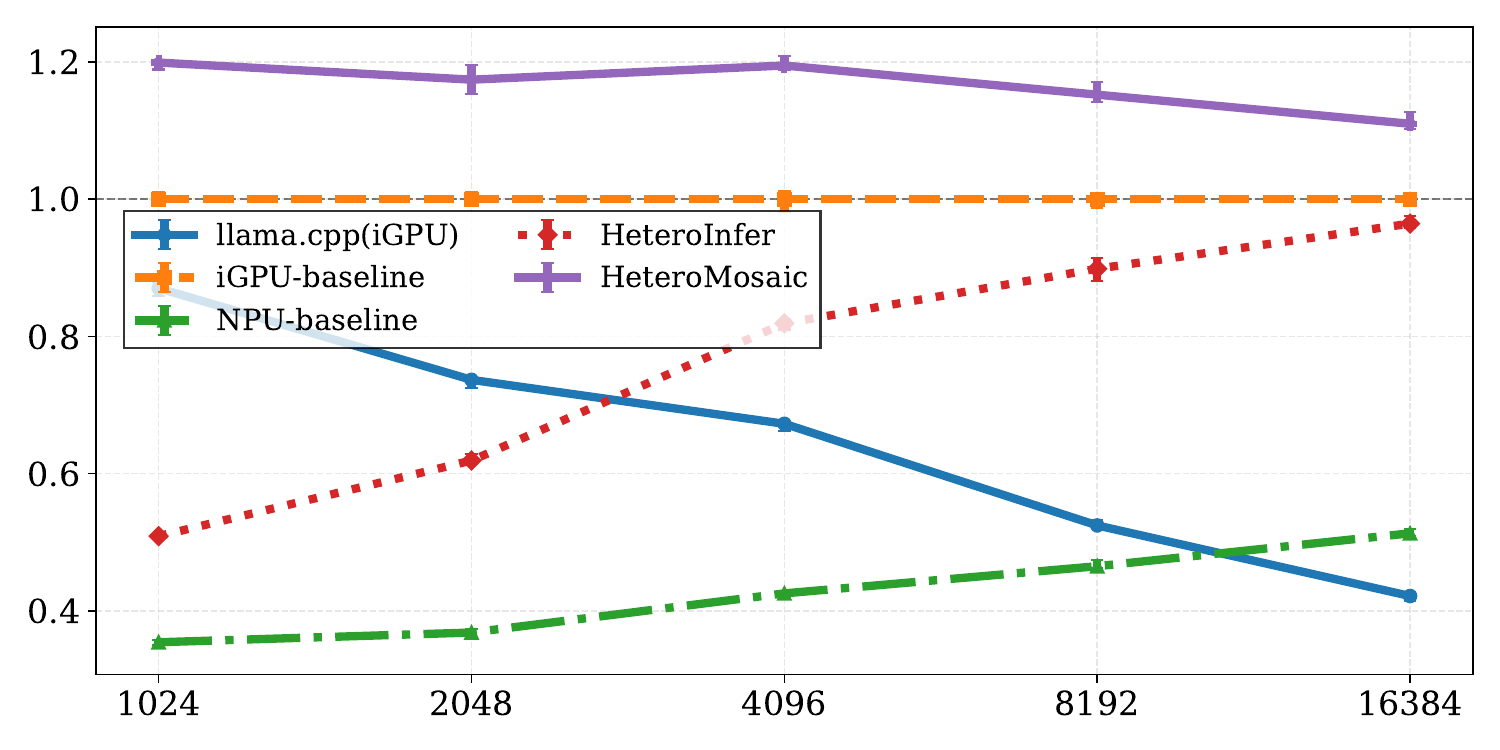} \\

        \modellabel{Llama3-8B} &
        \plotcell{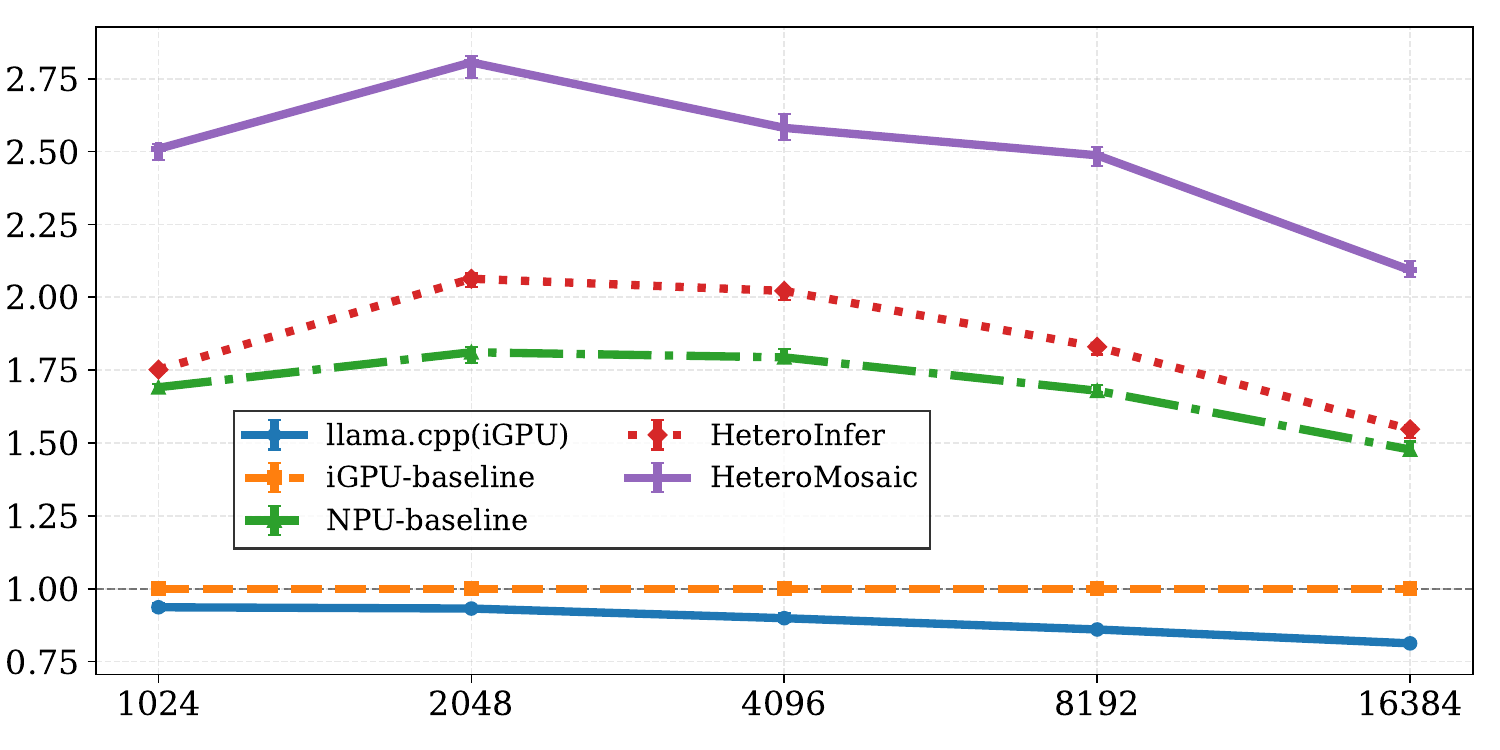} &
        \plotcell{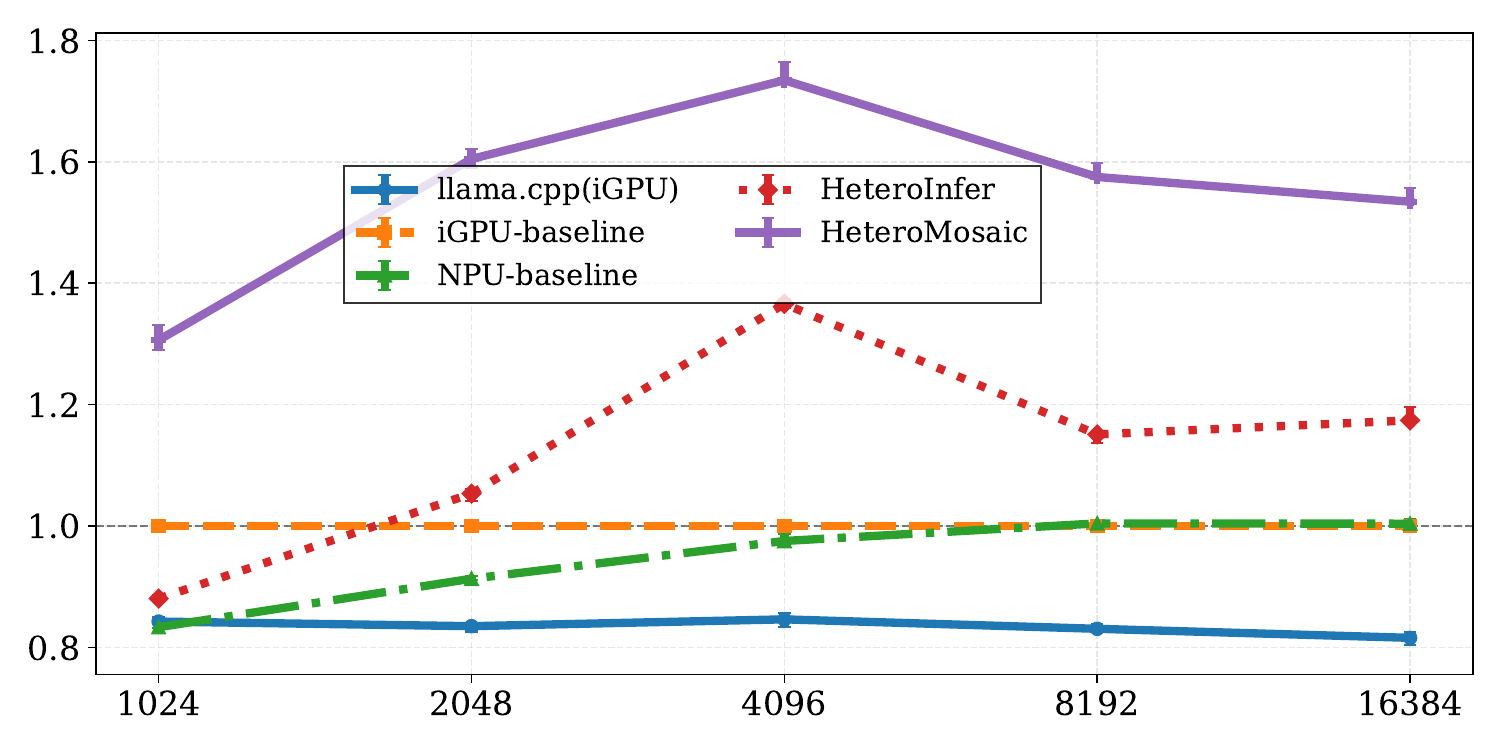} &
        \plotcell{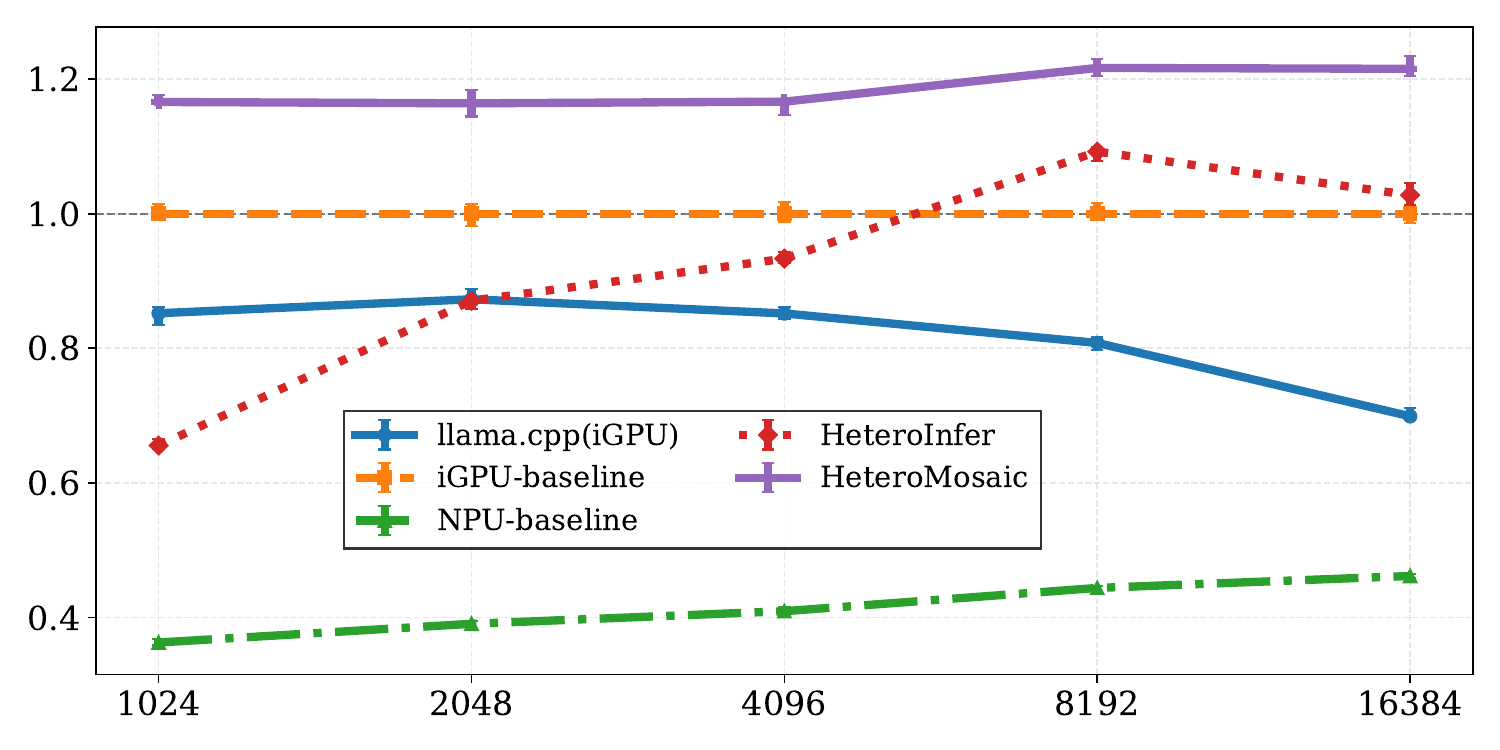} \\
        
        \modellabel{Qwen2.5-14B} &
        \plotcell{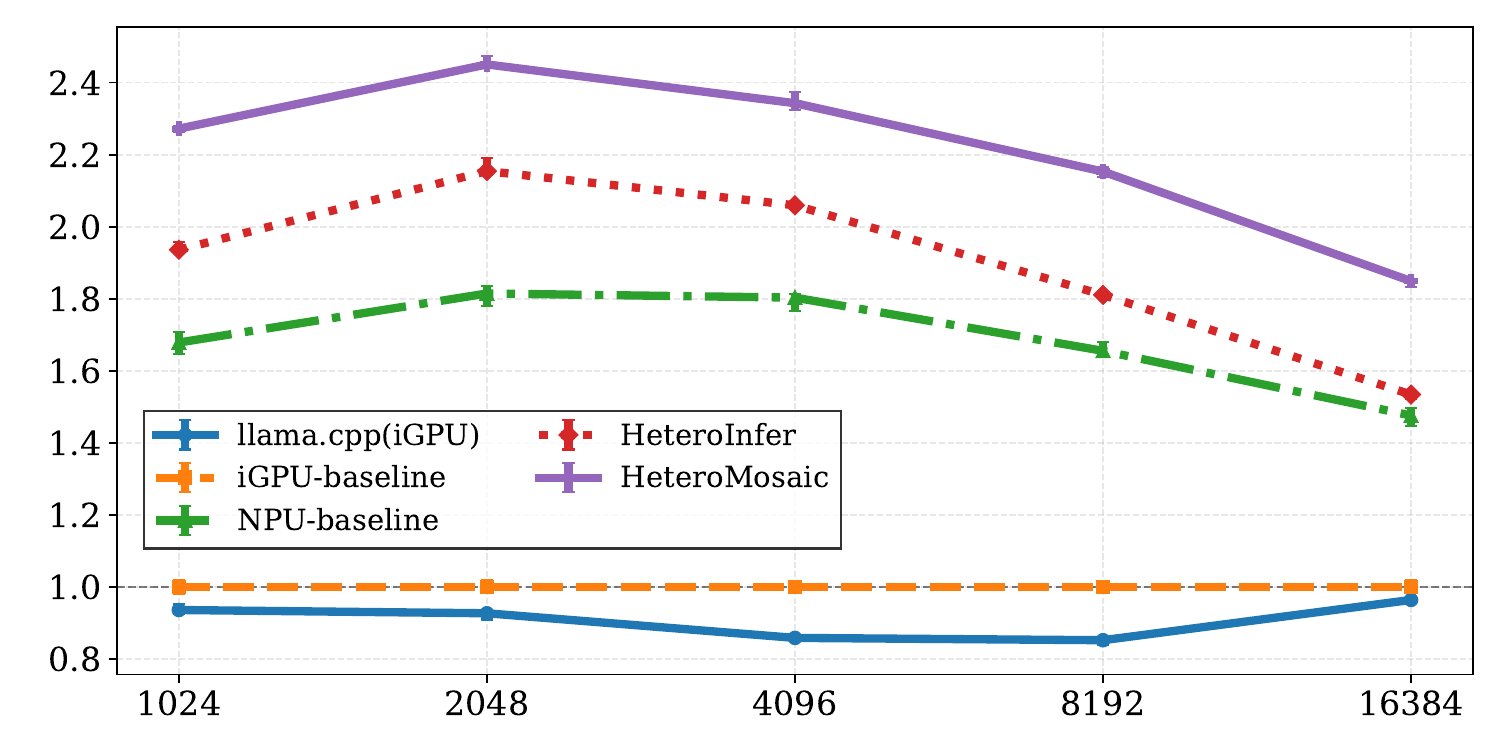} &
        \plotcell{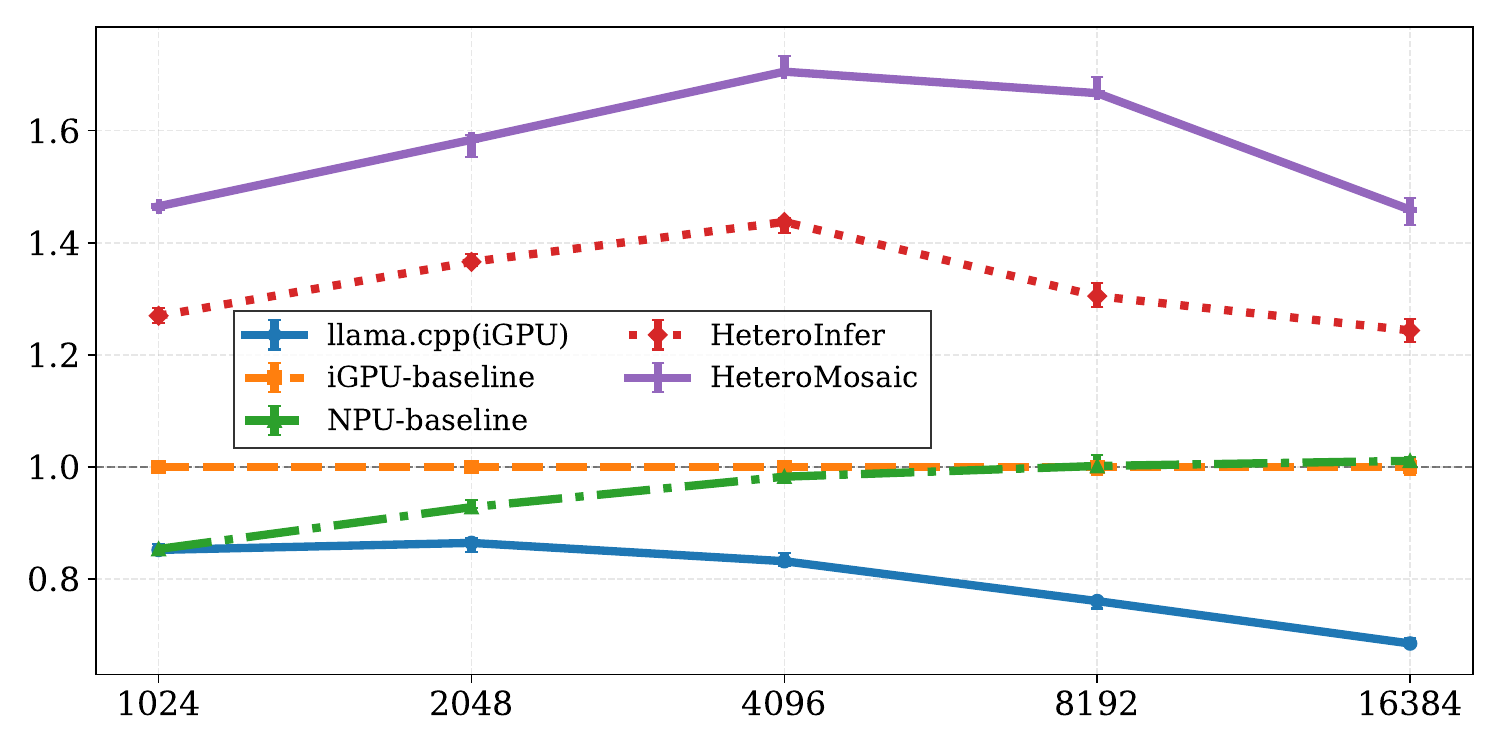} &
        \plotcell{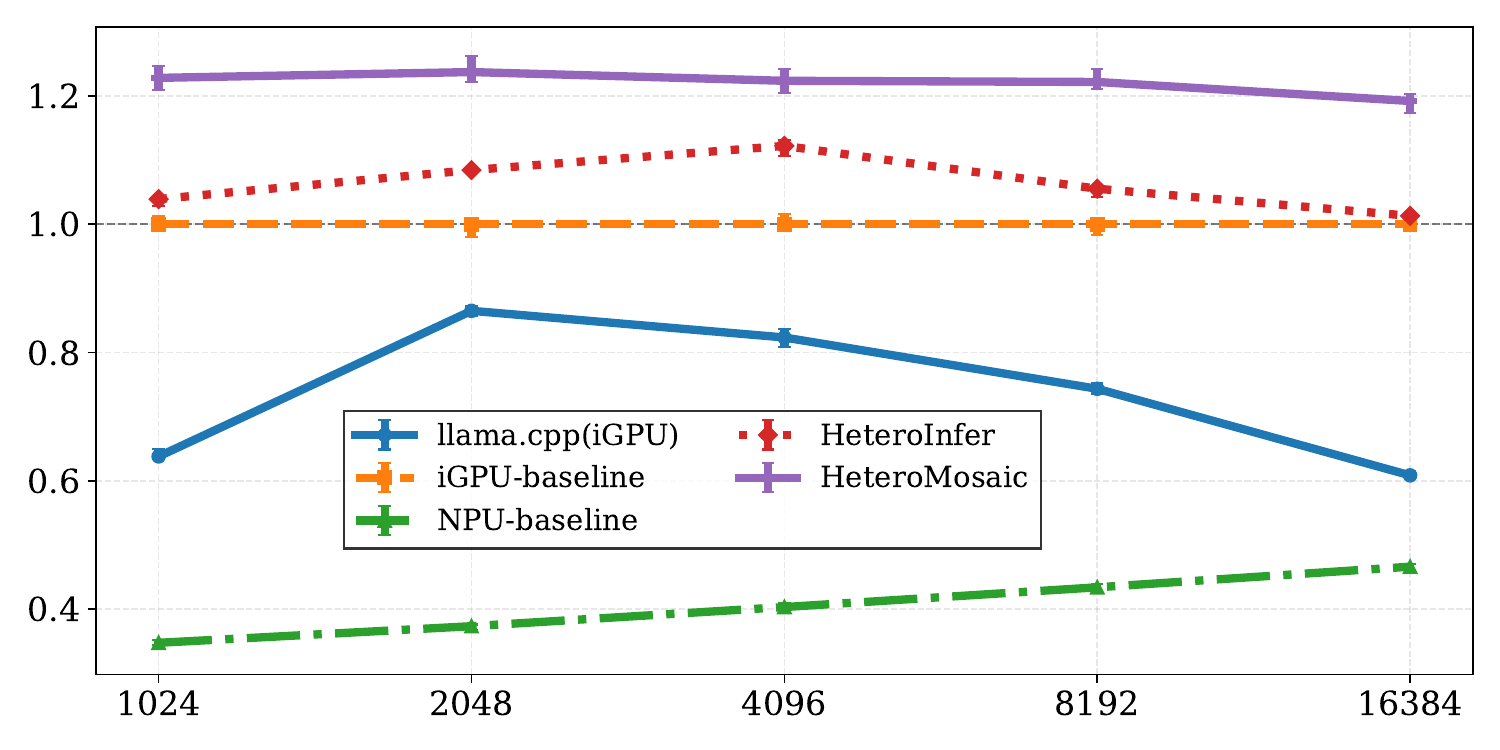} \\

        \modellabel{Llama3-70B} &
        \plotcell{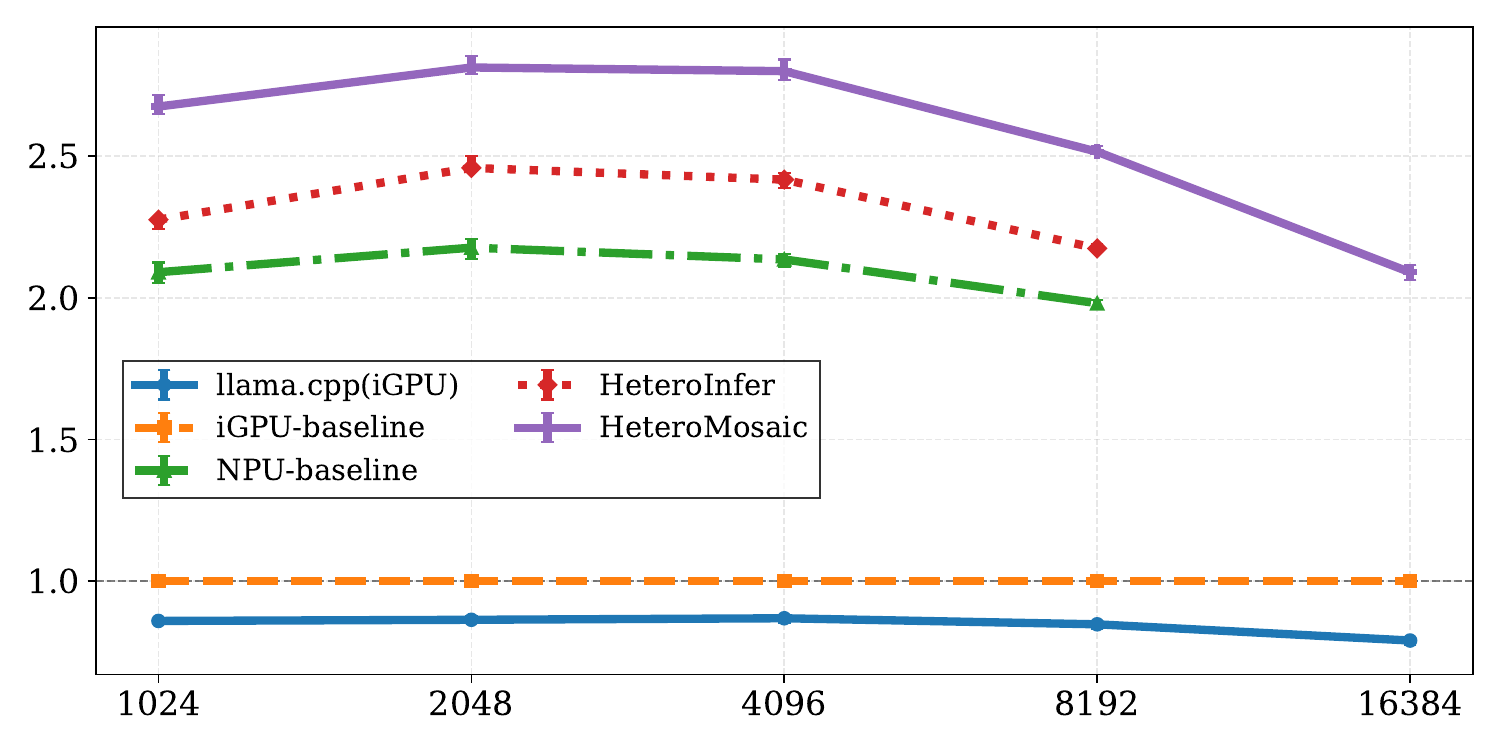} &
        \plotcell{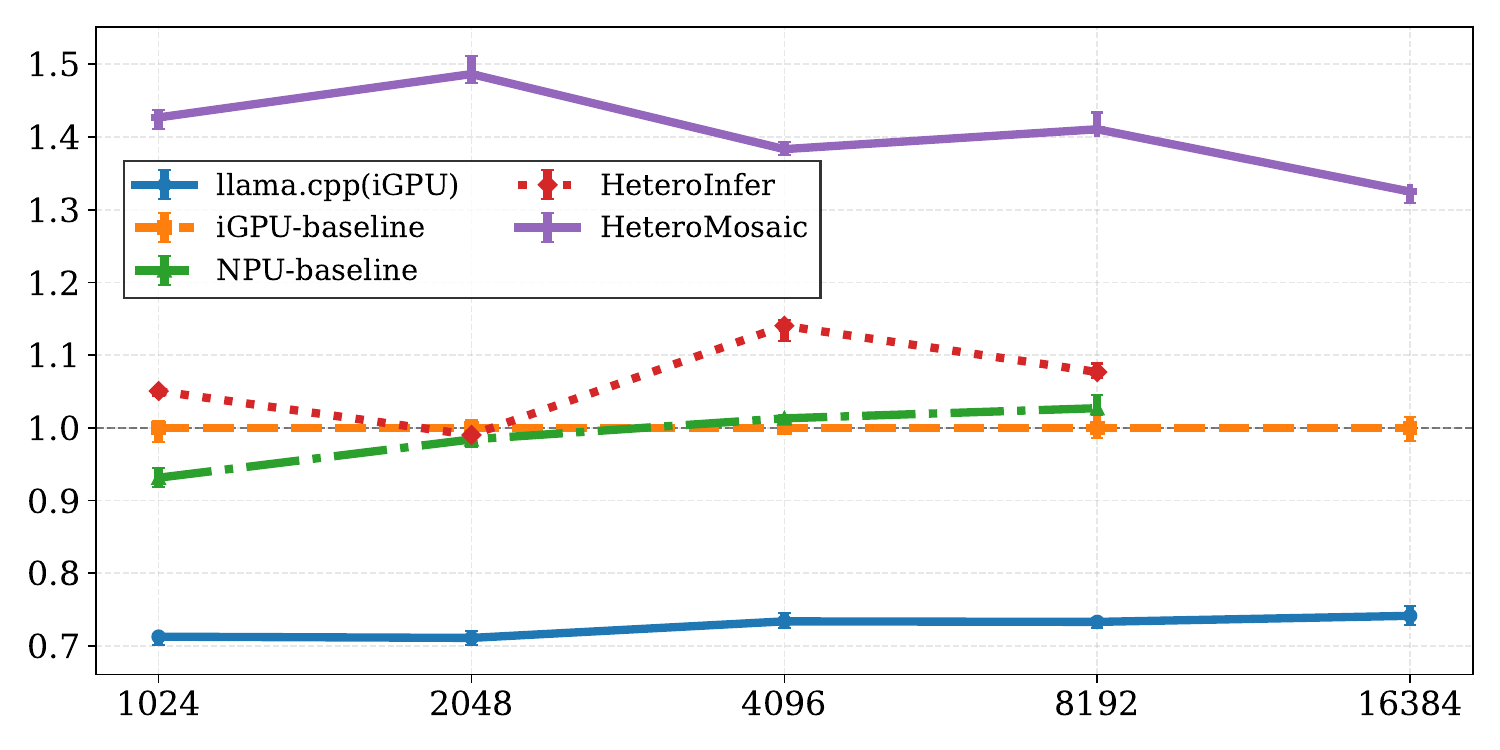} &
        \plotcell{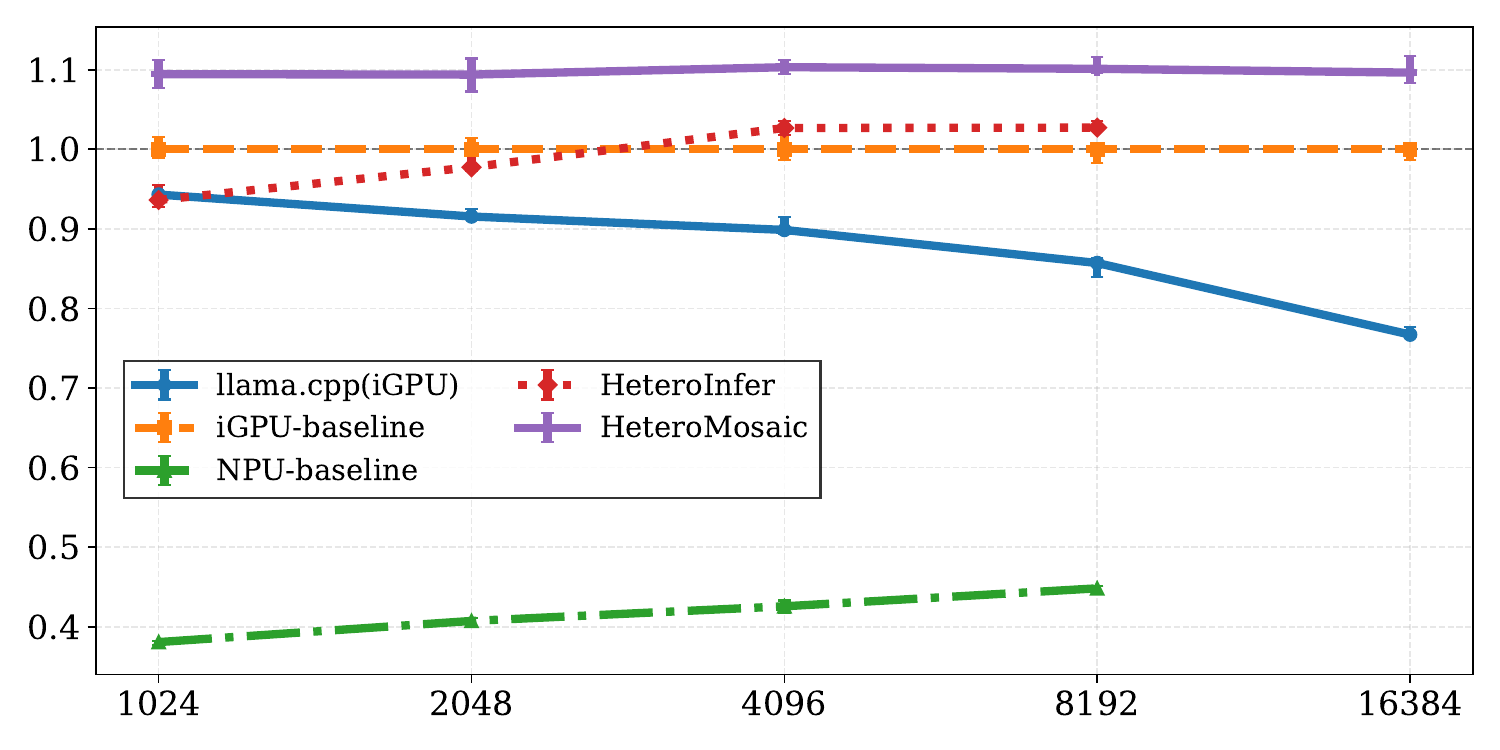} \\
    \end{tabular}

    \caption{End-to-end LLM prompt latency speedup across \Ryzen~ devices and across multiple AWQ W4A16 models from Hugging Face. Columns correspond to devices and rows to models. In each subplot, the x-axis shows input prompt length in tokens, and the y-axis shows speedup over the iGPU baseline, computed as baseline latency divided by method latency; higher is better. We note that for Llama3-70B, the 16K runs without micro-batching failed due to OOM.}
    \label{fig:end2end}
\end{figure*}

Thus, to evaluate whether and to what extent the heterogeneous gains observed in the microbenchmarks translate to the full end-to-end application, we evaluate HeteroMosaic across a series of LLMs with different model sizes and prompt lengths, as shown in Figure~\ref{fig:end2end}. This setting is substantially more challenging than the microbenchmark setting because full-model inference includes attention and KV-cache traffic, non-partitionable operators, layerwise shape variation, synchronization overhead, NPU runtime behavior, unified-memory contention, and DVFS/thermal effects. For each model--device pair, we report results across several prompt lengths and normalize performance to the corresponding iGPU baseline.

Our comparison includes five configurations: our iGPU-only baseline, our NPU-oriented baseline, \texttt{llama.cpp}~\cite{llama.cpp}\footnote{We base these results on \texttt{llama.cpp} commits from March 2026 and use the ROCm backend with \texttt{Q4\_K\_S}, the closest available proxy to AWQ-style W4A16 execution in \texttt{llama.cpp}. \texttt{Q4\_K\_S} is a \texttt{llama.cpp}-specific 4-bit K-quant format that keeps the linear-layer weights in 4-bit form, whereas \texttt{Q4\_K\_M} uses a mixed variant that retains selected tensors at higher precision. We therefore use \texttt{Q4\_K\_S} to better approximate a uniform 4-bit weight-only baseline.}, a HeteroInfer-style baseline, and HeteroMosaic.

For HeteroInfer~\cite{chen2025heteroinfer}, we implement a faithful re-creation based on the published design, since the original system is not open source and is not natively compatible with \Ryzen.\footnote{When details are ambiguous, we use conservative implementation choices that avoid penalizing the baseline. For example, \Ryzen~ NPUs exhibit less shape sensitivity than reported in the original paper, and Qualcomm's QNN stack differs from \Ryzen~ in both quantization support and kernel availability~\cite{scalingNPU,qualcomm_qnn_quant_2026,qualcomm_qnn_htp_op_package_2026}. We therefore use the strongest corresponding implementation available on our platform: fused IRON-based NPU kernels together with our heterogeneity-compatible iGPU AWQ W4A16 kernels, rather than directly mirroring QNN's channel-wise quantization path.} To validate that this re-creation captures the published behavior, we emulate the reported iGPU:NPU imbalance of roughly 1:10 by inserting bubbles on the iGPU path so that the effective ratio matches the performance characteristics described in HeteroInfer.\footnote{Under this calibration, on \Ryzen~ 7 350 with Llama3-8B at prompt length 256, our re-creation reaches roughly $6.1\times$ versus the reported $5.6\times$ relative to its respective iGPU baseline. We attribute the remaining gap to differences in device characteristics and implementation details not fully specified in the paper.}

Our iGPU baseline uses the same micro-batching schedule as \texttt{llama.cpp} to keep the comparison fair, yet it still consistently outperforms \texttt{llama.cpp}, largely because our iGPU kernels are more tightly optimized for \Ryzen. Our NPU-oriented baseline follows a HeteroInfer-like placement strategy, in which linear projections run on the NPU while operations such as attention and normalization remain on the iGPU.

\subsubsection{Analysis of Prefill Results}

Figure~\ref{fig:end2end} shows a clear device-level trend across all evaluated models. HeteroMosaic performs best on the left column (\Ryzen~ 7 350), remains clearly beneficial in the middle column (\Ryzen~ 9 HX 370), and still provides consistent gains in the right column (\Ryzen~ Max+ 395). This trend follows the underlying SoC balance point: as the system shifts from NPU-heavy to balanced and then to iGPU-heavy, the total heterogeneous headroom above the iGPU baseline shrinks. Averaged across the raw prefill results, HeteroMosaic achieves about $2.30\times$, $1.50\times$, and $1.17\times$ speedup over the iGPU baseline on the \Ryzen~ 7 350, \Ryzen~ 9 HX 370, and \Ryzen~ Max+ 395, respectively. This ordering is consistent with the roofline and GEMM microbenchmarks, but the absolute end-to-end gains are lower than the operation-level peaks because full inference also includes non-partitionable operators, attention and KV-cache bandwidth, synchronization overhead, finite split granularity, NPU reconfiguration, unified-memory contention, shape sensitivity, and DVFS/thermal effects. The HeteroInfer-style tensor-partitioning baseline falls even further below the GEMM microbenchmark peaks, showing that local operation partitioning alone does not preserve GEMM-level opportunity once embedded in the full execution graph.

A key point in Figure~\ref{fig:end2end} is that HeteroMosaic consistently outperforms both comparison strategies across all three devices. Relative to our NPU-oriented baseline, HeteroMosaic is still about $1.35\times$ faster on the \Ryzen~ 7 350, $1.60\times$ faster on the \Ryzen~ 9 HX 370, and $2.86\times$ faster on the \Ryzen~ Max+ 395 on average. This comparison makes clear that HeteroMosaic's advantage is not just a byproduct of favoring the NPU. Instead, these gains come from exposing additional graph-level overlap and then co-optimizing accelerator allocation around the actual critical path. HeteroMosaic also remains ahead of HeteroInfer across the full figure, with average improvements of about $1.22\times$, $1.28\times$, and $1.32\times$ on the same three devices. The rightmost column is especially instructive: even though the absolute iGPU-normalized headroom is smaller there, HeteroMosaic still retains an advantage because schedule-level overlap and latency shaping continue to matter after simple tensor placement has largely run out of room.

Across models, HeteroMosaic also shows that the amount of recoverable heterogeneous opportunity depends on model structure. Models with larger embedding dimensions, such as Qwen2.5-14B and Llama3-70B, naturally expose more operation-level heterogeneous work. Phi-3.5-3.8B is also relatively favorable because its MHA design retains larger $K$ and $V$ projections than the GQA-based models. Even so, the central pattern is consistent across rows: HeteroMosaic remains the strongest method because it combines operation-level partitioning with graph-level overlap and runtime-aware schedule shaping, rather than relying only on a placement policy for split-friendly dense layers. At long contexts, especially for Llama3-70B, the gains of all methods become more muted as execution moves deeper into the sustained-power regime, where the SoC has moved past short-lived turbo behavior and settles into a lower steady-state throughput under thermal and power limits. Even in this constrained regime, HeteroMosaic maintains the strongest overall performance. At 16K, it is the only heterogeneous method among these baselines that completes across all three devices while preserving positive speedup over the iGPU baseline. Beyond exposing heterogeneous overlap, its micro-batching also manages memory footprint, allowing large cases such as Llama3-70B at 16K to avoid OOM failures. Taken together, the prefill results show that tensor placement alone recovers only part of the available heterogeneous opportunity, whereas HeteroMosaic more fully realizes that opportunity by jointly reasoning about graph structure, operation shape, SoC balance, and runtime system effects.

\subsubsection{Analysis of Decode Results}
\begin{table}[!t]
\centering
\normalsize
\caption{Normalized TPS decode performance (HeteroMosaic / iGPU) across three \Ryzen~ platforms.}
\label{tab:normalized_results}
\scalebox{1}{
\begin{tabular}{@{} l c c c @{}}
\toprule
\textbf{Model} & 
\makecell[c]{\textbf{AMD Ryzen\textsuperscript{\texttrademark}}\\\textbf{AI 7 350}} &
\makecell[c]{\textbf{AMD Ryzen\textsuperscript{\texttrademark}}\\\textbf{AI 9 HX 370}} &
\makecell[c]{\textbf{AMD Ryzen\textsuperscript{\texttrademark}}\\\textbf{AI Max+ 395}} \\
\midrule
Phi3.5-3.8B & 1.00$\times$ & 1.11$\times$ & 1.11$\times$ \\
Llama3-8B   & 1.13$\times$ & 1.02$\times$ & 0.995$\times$ \\
Qwen2.5-14B & 1.07$\times$ & 0.99$\times$ & 0.996$\times$ \\
Llama3-70B  & 1.09$\times$ & 0.98$\times$ & 1.00$\times$ \\
\bottomrule
\end{tabular}
}
\end{table}

Decode performance is fundamentally memory-bound. However, similar to HeteroInfer~\cite{chen2025heteroinfer}, we find that heterogeneous GEMV can still be beneficial when the iGPU alone does not fully utilize the available memory system or when GEMV shapes underutilize the iGPU. As shown in Table~\ref{tab:normalized_results}, this occurs most consistently on the \Ryzen~ 7 350 for the larger evaluated models, where the weaker iGPU benefits modestly from additional CPU/NPU parallelism. On the \Ryzen~ 9 HX 370 and \Ryzen~ Max+ 395, heterogeneous decode is usually neutral or slightly harmful because the stronger iGPU already approaches the useful memory-bandwidth limit and the added synchronization overhead can outweigh the benefit. Phi3.5-3.8B is an exception: its smaller GEMV dimensions leave more room for shape- and launch-efficiency effects, allowing modest heterogeneous decode gains on the larger platforms as well. These results are consistent with the broader thesis of the paper: decode offers limited heterogeneous headroom on unified-memory SoCs once the iGPU can already drive sufficient memory traffic, whereas prefill remains the primary setting in which graph-level scheduling and latency shaping recover substantial heterogeneous benefit.

\subsection{Power Study}

\begin{figure}[t]
  \centering

  \newcommand{\trimplot}[1]{%
    \scalebox{1}[0.80]{%
      \raisebox{-0.5\height}{%
        \includegraphics[width=\linewidth,trim=12 12 10 10,clip]{#1}
      }%
    }%
  }

  \trimplot{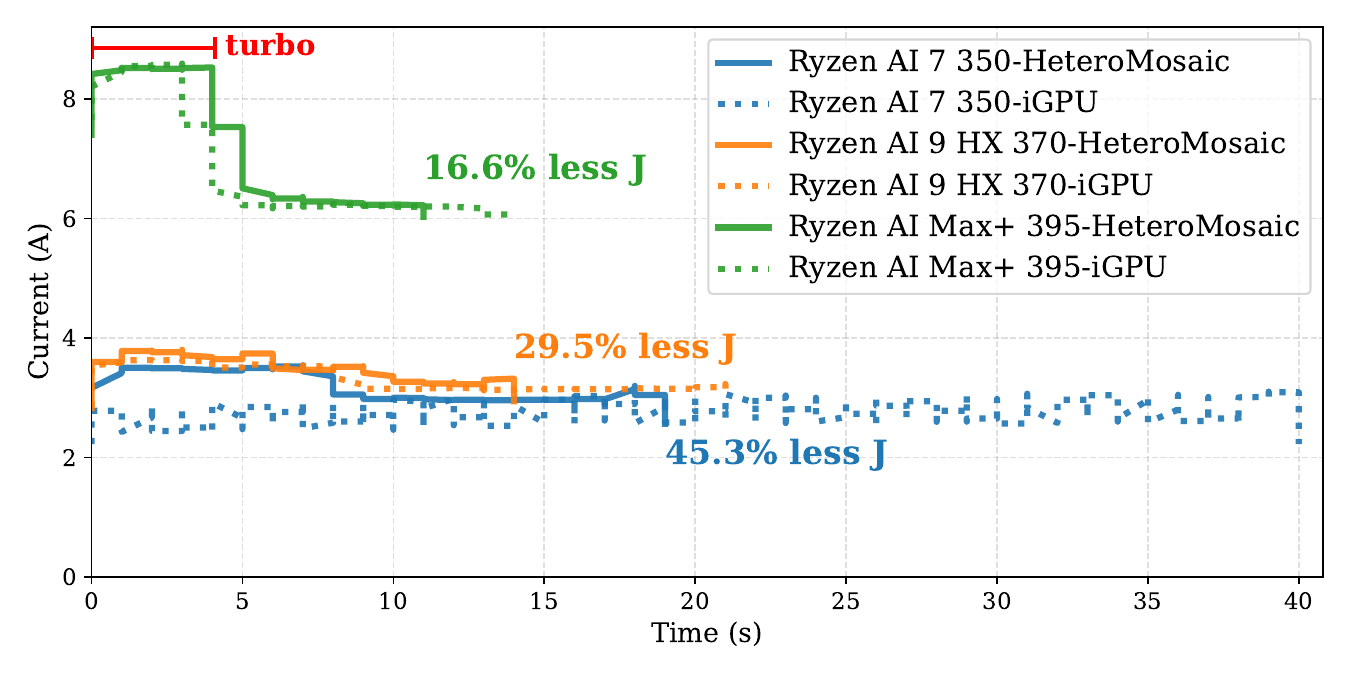}
  \caption{Measured current (A) over time during the same LLM inference workload across three platforms.}
  \label{fig:powerstudy}
\end{figure}

A natural concern with heterogeneous execution is that activating more accelerators will simply increase power. However, on TDP-constrained edge SoCs, as shown in Figure~\ref{fig:powerstudy}, the iGPU alone often already operates near the system power limit.\footnote{In heterogeneous execution, adding iGPU participation can cause the package to approach its TDP limit through iGPU DVFS regardless of the exact NPU fraction, so heavier NPU use does not necessarily translate into lower instantaneous power.} Instead, the more relevant question is whether heterogeneous scheduling can complete more useful work within roughly the same power envelope. Under this view, any speedup achieved without materially increasing peak power should reduce the total energy required to complete inference.

We validate this experimentally by measuring current using a lab bench power supply~\cite{owon_spe_series_dc_power_supply_2026} at a sampling rate of 10\,Hz. The results, summarized in Figure~\ref{fig:powerstudy}, support this intuition: HeteroMosaic improves performance without introducing a corresponding increase in peak current. We also observe that \Ryzen~ SoCs exhibit a turbo regime in which the chip operates at an elevated power level for a short interval before settling into a lower sustained operating point. HeteroMosaic can extend this high-performance interval, most clearly on the \Ryzen~ Max+ 395, which is consistent with our broader thesis that heterogeneity is most effective when it is co-optimized against real system behavior.

\subsection{Ablation Study}

\begin{figure}[t!]
  \centering

  \newcommand{\trimplot}[1]{%
    \scalebox{1}[0.80]{%
      \raisebox{-0.5\height}{%
        \includegraphics[width=\linewidth,trim=24 32 16 42,clip]{#1}
      }%
    }%
  }

  \trimplot{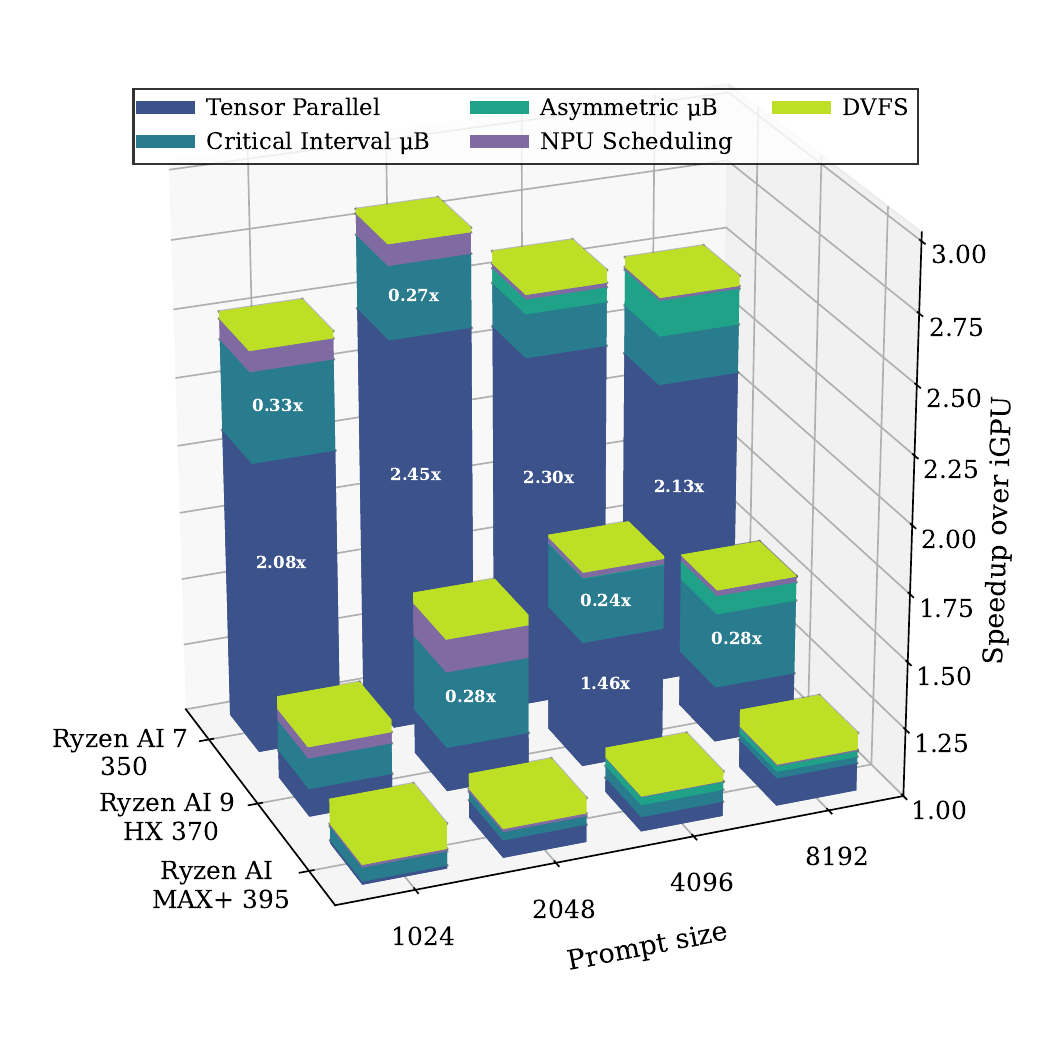}
  \caption{Ablation of HeteroMosaic for Llama3-8B across \Ryzen~ devices and prompt sizes. Bar height indicates speedup over the iGPU baseline, while colored segments show the contribution of each optimization.}
  \label{fig:ablation_study}
\end{figure}

An ablation study of HeteroMosaic is by nature non-trivial as its optimizations are co-dependent: removing one component changes the globally optimal schedule. In particular, tensor parallelism, asymmetric $\mu B$, DVFS-aware scheduling, critical-interval micro-batching, and NPU scheduling build on one another rather than contributing independently. For example, the full benefit of tensor parallelism cannot be realized to the same extent without critical-interval micro-batched scheduling, which exposes additional graph-level parallelism for heterogeneity to exploit. We therefore perform ablation by removing each component, re-co-optimizing the remaining system, and then measuring the resulting speedup so that each component’s contribution is isolated as fairly as possible.

Figure~\ref{fig:ablation_study} helps explain where HeteroMosaic’s gains come from. First, tensor parallelism provides the raw heterogeneous leverage by allowing compute to be mapped across devices; across systems and prompt lengths, this is the dominant source of speedup. Second, critical-interval scheduling determines whether that leverage survives globally, since locally beneficial mappings do not necessarily translate into lower end-to-end latency without schedule-level coordination. Third, as prompt length grows and attention becomes more dominant, the contribution of asymmetric $\mu B$ becomes more visible because it expands the schedule space and exposes additional schedulable work. Fourth, DVFS-aware scheduling matters because realized performance is not nominal performance: by partitioning work between the NPU and iGPU in a DVFS-aware manner, HeteroMosaic can keep the SoC in a higher turbo state for longer, increasing effective performance beyond what a static mapping would suggest. Finally, NPU scheduling matters because runtime effects such as reconfiguration overhead and redundant weight movement can otherwise erase a meaningful fraction of the gain.

\subsection{Comparison with Other Frameworks}

\begin{figure}[t]
  \centering

  \newcommand{\trimplot}[1]{%
    \scalebox{1}[0.80]{%
      \raisebox{-0.5\height}{%
        \includegraphics[width=\linewidth,trim=12 12 10 10,clip]{#1}%
      }%
    }%
  }

  \begin{subfigure}[t]{\linewidth}
    \centering
    \trimplot{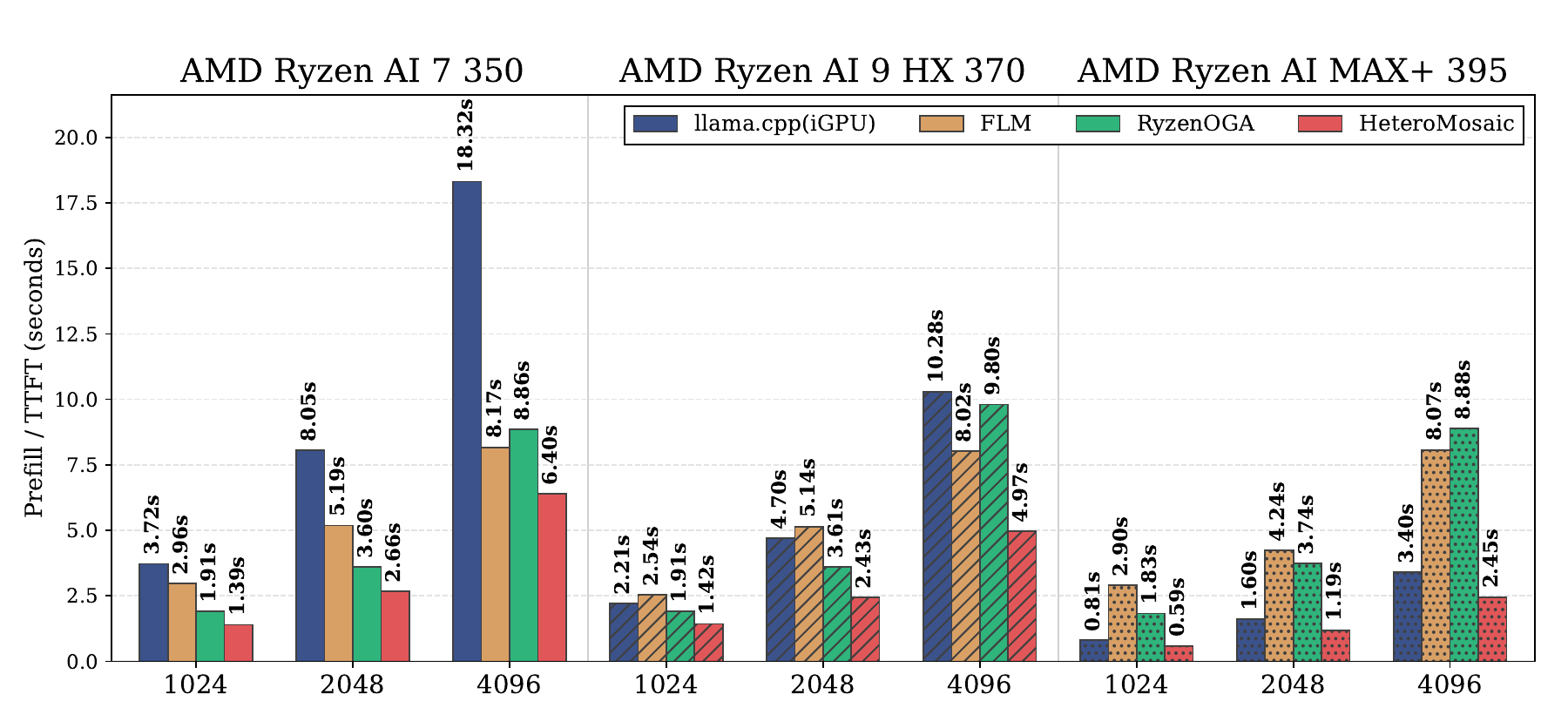}
    \caption{TTFT comparison. Lower is better.}
    \label{fig:crossplatform-ttft}
  \end{subfigure}

  \begin{subfigure}[t]{\linewidth}
    \centering
    \trimplot{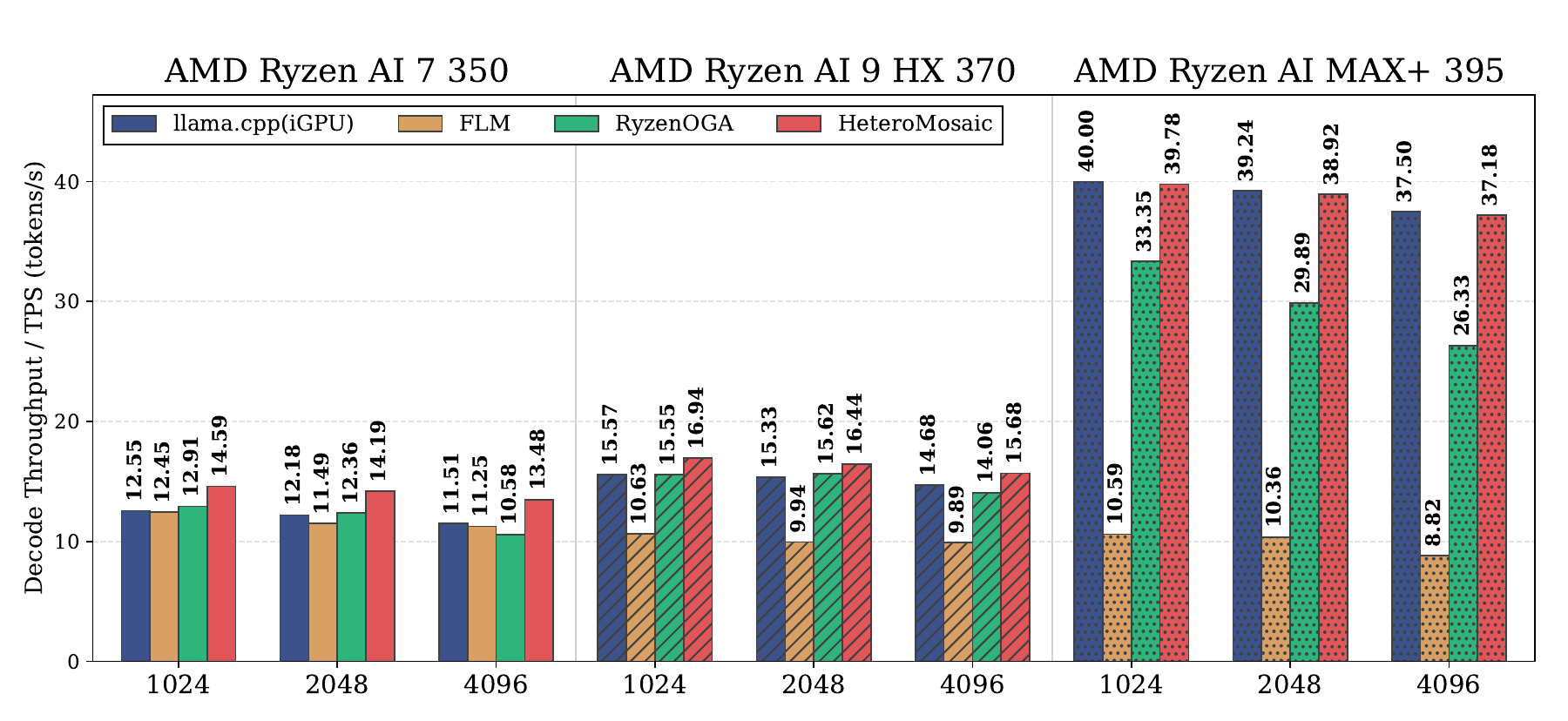}
    \caption{TPS comparison. Higher is better.}
    \label{fig:crossplatform-tps}
  \end{subfigure}

  \caption{Cross-framework comparison for Llama3-8B across \Ryzen~ platforms.}
  \label{fig:crossplatform}
\end{figure}

To contextualize HeteroMosaic against \texttt{llama.cpp} iGPU-ROCm~\cite{llama.cpp} and production AMD LLM frameworks, including FastFlowLM (FLM)~\cite{fastflowlm_github_2026} and the Ryzen ONNX Hybrid Runtime (Ryzen-OGA)\footnote{Ryzen-OGA runs prompt encoding on the NPU and decode on the iGPU.}~\cite{amd_oga_2026}, we evaluate Llama3-8B~\cite{llama} across \Ryzen~ platforms. We report raw TTFT and TPS in Figure~\ref{fig:crossplatform} so that cross-platform and cross-device performance can be compared more directly. These comparisons are not numerically identical across implementations: for example, \texttt{llama.cpp} on the iGPU uses integer computation, Ryzen-OGA uses BFP16~\cite{song2018computation} activations, and HeteroMosaic uses BF16, while the exact internal behavior of FLM is not openly documented. Therefore, these experiments should be interpreted as a coarse cross-framework and cross-platform comparison rather than a strict apples-to-apples evaluation.

For TTFT, HeteroMosaic is consistently the strongest solution across the evaluated platforms, with the largest benefits appearing when heterogeneous resources can be effectively exposed and scheduled at the graph level. A particularly notable result is that HeteroMosaic on the \Ryzen~ 7 350 achieves a 2048-token TTFT of 2.66\,s, which is not only substantially faster than the same-platform \texttt{llama.cpp} baseline of 8.05\,s, but also faster than the stronger llama.cpp iGPU baseline on the AMD Ryzen™ AI 9 HX 370 of 4.70\,s. This corresponds to a 1.76$\times$ TTFT advantage over the \Ryzen~ 9 HX 370 \texttt{llama.cpp} baseline, demonstrating that well-orchestrated heterogeneous execution can outperform stronger iGPU-only baselines. This suggests that balanced heterogeneous compute, low-overhead shared memory, and fine-grained synchronization can provide more practical benefit for transformer inference than simply increasing iGPU resources alone.

For TPS, the picture is more nuanced because decode is more strongly constrained by memory bandwidth. HeteroMosaic remains competitive and often leads, but the gains are smaller than for TTFT, especially on the \Ryzen~ Max+ 395, where both HeteroMosaic and \texttt{llama.cpp} already operate near the bandwidth ceiling. In contrast, on the \Ryzen~ 7 350 and \Ryzen~ 9 HX 370, HeteroMosaic continues to provide measurable throughput advantages over FLM and Ryzen-OGA while remaining competitive with, or exceeding, the \texttt{llama.cpp} baseline.

\section{Discussion and Future Work}
\label{sec:discussion}

Although the evaluation in this paper uses \Ryzen~ as a concrete implementation target, the broader goal is to provide a principled methodology to reason about when fine-grained heterogeneity becomes useful. In its current realization, HeteroMosaic benefits from three platform properties: programmable accelerators, shared or low-cost memory access across devices, and low-overhead synchronization mechanisms. These properties are not unique to \Ryzen, but they are also not guaranteed across all heterogeneous systems. Thus, future systems that build on the insights of HeteroMosaic would need to develop the corresponding platform-specific runtime mechanisms, including GPU/NPU kernels, memory-layout compatibility, queue management, and synchronization, while preserving the same scheduling abstraction.

Beyond LLM inference, the same scheduling formulation should be applicable to other transformer-based workloads, including vision-language-action models for robotics such as OpenPI/\(\pi_0\), \(\pi_{0.5}\), and NVIDIA GR00T~\cite{physicalintelligence2025openpi,physicalintelligence2025openpi_blog,black2024pi0,physicalintelligence2025pi05,nvidia2025gr00tn1,nvidia2025gr00tproduct}, as well as transformer-based diffusion and flow models~\cite{bao2023uvit,peebles2023dit,esser2024scaling,flux2024}. These workloads contain fundamentally similar mixtures of dense projections, attention-like stages, memory-sensitive operators, and cross-stage dependencies, making them natural candidates for graph-level heterogeneous scheduling. HeteroMosaic, when applied to these similar workloads, may expose additional opportunities because their modality-specific components can create more heterogeneous scheduling choices than text-only LLMs. However, these workloads also introduce new constraints, such as image-token preprocessing, vision encoder stages, control-loop latency requirements, and diffusion-step dependencies. Thus, extending HeteroMosaic to these settings would require extending the current model from a single transformer execution path to multiple heterogeneous model stages that connect and execute together, while also accounting for how modality-specific stages interact with the critical path.

The cross-platform TTFT results also suggest a hardware-design implication where a smaller heterogeneous platform may have the potential to outperform a stronger iGPU-only baseline if the software can expose enough parallel work and coordinate the accelerators efficiently. This does not imply that larger iGPUs are unhelpful; rather, it suggests that balanced accelerator design, shared-memory efficiency, and low-overhead synchronization can matter as much as raw single-accelerator scale for transformer inference.

Finally, extending this formulation to heterogeneous datacenter systems is another possible direction, where the same heterogeneous scheduling idea could be applied across a mixture of accelerators spanning different CPU and GPU generations, compute capabilities, and more unconventional accelerators such as field-programmable gate arrays (FPGAs). However, this extension would require explicit communication modeling. Unlike unified-memory edge SoCs, datacenter accelerators often communicate through PCIe~\cite{pcisig_pcie_base_7_0}, CXL~\cite{cxl_consortium_spec}, NVLink~\cite{nvidia_nvlink}, or network-level links. In that setting, accelerator placement must account for activation movement, synchronization latency, collective overheads, and interconnect contention. The same critical-interval optimization principle may still apply, but the cost model would first need to treat communication as a first-class scheduling constraint and then apply heterogeneous scheduling around that constraint.

\section{Conclusion}

Edge AI is increasingly defined by heterogeneous SoCs, yet current inference systems still treat heterogeneity too coarsely and therefore underutilize the full combination of on-chip compute resources. In this paper, we present HeteroMosaic, a heterogeneity-first framework that treats edge LLM inference primarily as a scheduling problem rather than merely a placement problem. HeteroMosaic first uses a heterogeneous roofline model to determine when heterogeneity should help in principle, then restructures execution through causal micro-batching to expose latent cross-device overlap, and finally realizes that opportunity through latency shaping and trace-guided critical-interval co-optimization under real system effects.

Across three \Ryzen~ platforms, HeteroMosaic improves end-to-end LLM inference over strong baselines without increasing peak power. This shows that, on TDP-constrained edge SoCs, co-optimized heterogeneity can complete more work within the same practical power envelope, improving both energy efficiency and tokens per watt. Although the exact implementation depends on platform-specific runtime mechanisms, the underlying scheduling principles are broader than any one SoC family. Taken together, HeteroMosaic shows that fine-grained heterogeneity, when exposed and controlled through scheduling, is a practical path toward faster and more energy-efficient edge LLM inference.

\section*{Acknowledgment}
We thank Paul Hartke from AMD, and other AMD colleagues who provided valuable feedback and technical guidance throughout this work. This work is supported in part by the AMD Center of Excellence at UIUC.

\subsection*{Code Availability} The source code for HeteroMosaic is available at \url{https://github.com/UIUC-ChenLab/heteroMosaic}.


\bibliographystyle{ACM-Reference-Format}
\bibliography{papers.bib}

\end{document}